\documentclass[longauth]{aa}
\usepackage[varg]{txfonts}

\usepackage{natbib}
\bibpunct{(}{)}{;}{a}{}{,} 
\usepackage{graphicx}	
\usepackage{amsmath}	
\usepackage{placeins}
\usepackage[hyperfootnotes=false, linktocpage=true, breaklinks=true, colorlinks=true, linkcolor=blue, citecolor=blue, urlcolor=blue]{hyperref}
\usepackage{xcolor}

\usepackage{longtable}
\usepackage{xcolor}

\usepackage{booktabs}

\newcommand{\funit}{~erg cm$^{-2}$ s$^{-1}$} 
\newcommand{\ergs}{~erg s$^{-1}$}

\newcommand{\cmsq}{~cm$^{-2}$ }   
\newcommand{\nh}{$\rm{N_{H}}$}
\newcommand{\chisq}{$\rm{\chi^{2}}$ }
\newcommand{\chisqr}{$\rm{\chi_{\nu}^{2}}$ }

\def\chandra{{\it Chandra}}

\def\gaia{{\it Gaia}}

\begin{document}

\title{EWOCS-IV: 1\,Ms ACIS Chandra observation of the supergiant B[e] star Wd1-9}

\author{K. Anastasopoulou\inst{1,2}\thanks{E-mail: konstantina.anastasopoulou@cfa.harvard.edu}
\and  M. G. Guarcello\inst{2}
\and J. J. Drake\inst{3}
\and B. Ritchie\inst{4}
\and M. De Becker\inst{5}
\and A. Bayo\inst{6}
\and  F. Najarro\inst{7}
\and I. Negueruela\inst{8}
\and S. Sciortino\inst{2}
\and E. Flaccomio\inst{2}
\and R. Castellanos\inst{7}
\and J. F. Albacete-Colombo\inst{9} 
\and M. Andersen\inst{6}
\and F. Damiani\inst{2}
\and F. Fraschetti\inst{1}
\and M. Gennaro\inst{10,11}
\and S. J. Gunderson\inst{12}
\and C. J. K. Larkin\inst{13,14,15}
\and J. Mackey\inst{16}
\and A. F. J. Moffat\inst{17}
\and P. Pradhan\inst{18}
\and S. Saracino\inst{19,20}
\and I. R. Stevens\inst{21}
\and G. Weigelt\inst{22}
}

\institute{Center for Astrophysics $|$ Harvard \& Smithsonian, 60 Garden Street, Cambridge, MA 02138, USA
   \and Istituto Nazionale di Astrofisica (INAF) – Osservatorio Astronomico di Palermo, Piazza del Parlamento 1, 90134 Palermo, Italy   
   \and Lockheed Martin Solar and Astrophysics Laboratory, 3251 Hanover Street, Palo Alto, CA 94304, USA  
  \and  School of Physical Sciences, The Open University, Walton Hall, Milton Keynes MK7 6AA, UK 
  \and Space Sciences, Technologies and Astrophysics Research (STAR) Institute, University of Liège, Quartier Agora, 19c, Allée du 6 Aôut, B5c, B-4000 Sart Tilman, Belgium 
\and  European Southern Observatory, Karl-Schwarzschild-Strasse 2, D-85748 Garching bei M\"{u}nchen, Germany 
\and   Departamento de Astrofísica, Centro de Astrobiología, (CSIC-INTA), Ctra. Torrejón a Ajalvir, km 4, Torrejón de Ardoz, E-28850 Madrid, Spain
\and
        Departamento de Física Aplicada, Facultad de Ciencias, Universidad de Alicante, Carretera de San Vicente s/n, E-03690, San Vicente del Raspeig, Spain 
 \and Universidad de Rio Negro, Sede Atlántica - CONICET, Viedma CP8500, Río Negro, Argentina 
  \and
        Space Telescope Science Institute, 3700 San Martin Dr, Baltimore, MD, 21218, USA 
         \and
        The William H. Miller III Department of Physics \& Astronomy, Bloomberg Center for Physics and Astronomy, Johns Hopkins University, 3400 N. Charles Street, Baltimore, MD 21218, USA 
\and Kavli Institute for Astrophysics and Space Research, Massachusetts Institute of Technology, 77 Massachusetts Ave., Cambridge, MA 02139, USA
\and Max-Planck-Institut f\"{u}r Kernphysik, Saupfercheckweg 1, D-69117 Heidelberg, Germany
\and Astronomisches Rechen-Institut, Zentrum f\"{u}r Astronomie der Universit\"{a}t Heidelberg, M\"{o}nchhofstr. 12-14, D-69120 Heidelberg, Germany
\and Max-Planck-Institut f\"{u}r Astronomie, K\"{o}nigstuhl 17, D-69117 Heidelberg, Germany
 \and Astronomy \& Astrophysics Section, School of Cosmic Physics, Dublin Institute for Advanced Studies, DIAS Dunsink Observatory, Dublin D15 XR2R, Ireland
 \and  Universite de Montreal, Montreal, Quebec, H2V 0B3, Canada
\and Embry Riddle Aeronautical University, Department of Physics \& Astronomy, 3700 Willow Creek Road, Prescott, AZ 86301, USA  
\and Istituto Nazionale di Astrofisica (INAF) – Osservatorio Astrofisico di Arcetri, Largo E.Fermi 5, 50125 Firenze, Italy
\and Astrophysics Research Institute, Liverpool John Moores University, 146 Brownlow Hill, Liverpool L3 5RF, UK
 \and School of Physics \& Astronomy, University of Birmingham, Birmingham B15 2TT, UK
\and Max Planck Institute for Radio Astronomy, Auf dem Hügel 69, D-53121 Bonn, Germany }

\date{Received xxxx / Accepted xxxx}
 \abstract
   {Supergiant B[e] (sgB[e]) stars are exceptionally rare objects, with only a handful of confirmed examples in the Milky Way. The evolutionary pathways leading to the sgB[e] phase remain largely debated, highlighting the need for additional observations. The sgB[e] star Wd1-9, located in the massive cluster Westerlund 1 (Wd1), is enshrouded in a dusty cocoon—likely the result of past eruptive activity—leaving its true nature enigmatic.}  
   {We present the most detailed X-ray study of Wd1-9 to date, using X-rays that pierce through its cocoon with the aim to uncover its nature and evolutionary state.}
{We utilize 36 \chandra{} observations of Wd1 from the "Extended Westerlund 1 and 2 Open Clusters Survey" (EWOCS), plus eight archival datasets, totalling 1.1 Ms. This dataset allows investigation of long-term variability and periodicity in Wd1-9, while X-ray colours and spectra are analysed over time to uncover patterns that shed light on its nature.
}
 {Wd1-9 exhibits significant long-term X-ray variability, within which we identify a strong $\sim$14-day periodic signal. We interpret this as the orbital period, marking the first period determination for the system.
 The X-ray spectrum of Wd1-9 is thermal and hard ($kT \sim 3.0$ keV), resembling the spectra of bright Wolf-Rayet (WR) binaries in Wd1, while a strong Fe emission line at 6.7 keV indicates hot plasma from a colliding-wind X-ray binary.}
  {Wd1-9, with evidence of past mass loss, circumbinary material, a hard X-ray spectrum, and a newly detected 14-day period, displays all the hallmarks of a binary—likely a WR+OB—that recently underwent early Case B mass transfer. Its sgB[e] classification is likely phenomenological reflecting emission from the dense circumbinary material. This places Wd1-9 in a rarely observed phase, possibly revealing a newly formed WN star, bridging the gap between immediate precursors and later evolutionary stages in Wd1.}

\authorrunning{K. Anastasopoulou et al.}
\titlerunning{1\,Ms \chandra{} observation of Wd1-9} 
\maketitle




\section{Introduction}

Supergiant B[e] (sgB[e]) stars are a rare and enigmatic class of evolved massive stars characterised by their 
strong Balmer emission lines, forbidden line emission
from low-ionization metals, and significant infrared excess revealing dense, dusty circumstellar environments \citep[for a review see][]{kraus19}.
The exact evolutionary state of sgB[e] stars is still debated and is quite uncertain, but they are thought to represent transitional phases in the evolution of massive stars. In the Milky Way, only about a dozen confirmed examples of sgB[e] stars have been identified, making them a uniquely intriguing population to study. 
Most sgB[e] stars are not found in stellar clusters, complicating efforts to age date them and place them in a well-defined evolutionary context. Wd1-9 \citep{clark05,clark13}, in the supermassive star cluster Westerlund 1 (Wd1), is one of only two such stars identified within Galactic star clusters, the other being MWC~137 in Sh2-266 \citep[see][]{mehner16}. Wd1 is rich in evolved massive stars and offers a unique opportunity to study this transient evolutionary phase alongside other massive stars formed in the same environment and at the same time.

Wd1-9 \citep[referred to as Ara C in][]{borgman70} was initially classified as a B[e] star with an extended shell \citep{westerlund87}, and later on it was found by \citet{clark05} to fulfil the classification criteria for  sgB[e] stars of \citet{lamers98}. 
Optical observations revealed that the source displays a rich emission-line spectrum, with no detectable photospheric features \citep{clark05,clark13} and non-periodic photometric variability \citep{bonanos07}. In the mid-infrared, Wd1-9 exhibits strong emission, while the emission lines and shape of an Infrared Space Observatory (ISO) Short Wavelength Spectrometer (SWS) spectrum implied a shock-heated gas with a temperature higher than $\sim$80\,kK, and equatorially concentrated silicate dust, with a mass of
$\sim$$10^{-4}$ $M_{\odot}$ \citep{clark98,clark13}. Moreover, JWST Mid-Infrared Instrument (MIRI) observations revealed the existence of an asymmetric outflow \citep{guarcello25}.

Radio observations \citep{clark98,dougherty05,dougherty10,fenech17,fenech18,andrews19} identified Wd1-9 as the brightest radio source in Wd1, characterised by a compact, intense thermal emitter surrounded by an extended nebula reminiscent of luminous blue variables (LBVs). It stands among the most luminous radio thermal-emitting stars known, with a current mass-loss rate of $\sim$9$\times$10$^{-5}$$M_{\odot}$yr$^{-1}$, nearly an order of magnitude higher than other cluster members. The nebula likely originated from an earlier wind epoch, with a density about three times greater than the current stellar wind \citep{dougherty10}. Such past explosive phases, as seen in the radio data, are reminiscent of those seen in cool hypergiants, LBVs, or during rapid Roche-lobe overflow in interacting binaries \citep[e.g.][]{lobel03, langer03,groh10}.
At X-ray wave, approximately 60\,ks of \chandra{} observations have revealed a hard thermal X-ray spectrum and a high luminosity, consistent with a colliding-wind X-ray binary scenario \citep{skinner06,clark08}.

Although Wd1-9 has been studied at many wavelengths its exact nature remains a mystery.
\citet{clark13} proposed that the emission line spectrum likely arises in part from a compact circumstellar envelope rather than exclusively from a stellar wind, and the system is best described as a short-period massive binary that has recently undergone Case~A mass transfer (mass transfer that occurs while the donor is still burning hydrogen in its core), encircled by a substantial dusty disk. 
Thus far, the only direct evidence for binarity comes from the X-ray data, with the luminosity and hardness incompatible with expectations for an isolated star, and despite the observed optical variability, no orbital period has been detected for this system.

In this work, we analyse the deepest X-ray observation available (1\,Ms ACIS-I \chandra) from the Extended Westerlund 1 and 2 Open Clusters Survey (EWOCS).
These data offer a unique opportunity to examine the X-ray properties of Wd1-9 in detail, search for potential periodic signals, and contribute to unravelling the nature of this enigmatic source.
This is the fourth paper in the EWOCS series and is organised as follows. Section \ref{dataandmethods} provides an overview of the data and methods used thorough this work. In Section \ref{analysisandresults}, we present our results, including investigations of variability and periodicity, X-ray colours, and spectral analysis. Section \ref{discussion} presents a discussion of our findings, and Section \ref{conclusions} our conclusions.
Throughout this work, we adopt a distance to Wd1 of 4.23\,kpc \citep{negueruela22}.

\section{Data and Methods}\label{dataandmethods}

In the following sections, we provide a description of the data and outline the analysis methods employed throughout this study. A more detailed description of our methodology can be found in the second paper of the EWOCS series \citep{konna24}.

\subsection{Observations and data reduction}\label{observations}

For this study, we use 44 \chandra{} observations of Wd1, 36 from the EWOCS \chandra{} Large Program (Proposal 21200267, PI Guarcello; ACIS-I; 967.80 ks), and eight archival, pre-EWOCS observations (PIs Skinner, Israel, Rea; ACIS-S; 151.93 ks). EWOCS data target Wd1, while two archival observations are aimed at Wd1 and six at the magnetar CXO J164710.20-455217. EWOCS observations were conducted from June 2020 to August 2021; pre-EWOCS observations span June 2005 to February 2018. Further details on observations, technical specifications, and data processing are available in \citet{guarcello24}, and a list of observations is in the Chandra Data Collection (CDC) 153\footnote{\href{https://doi.org/10.25574/cdc.153}{https://doi.org/10.25574/cdc.153}}.

\subsection{Variability and periodicity investigation}\label{varmethods}

The EWOCS dataset spans 14 months, providing a unique opportunity to identify long-term flux modulations.
To investigate long-term variability and periodicity, we utilised broad-band (0.5--8\,keV) probability-weighted light curves from the individual observations that we produced using the \textit{glvary} tool of CIAO, an implementation of the Gregory-Loredo algorithm variability test \citep{gl92}. 
This approach allows for a more detailed analysis of long-term trends and potential periodic signals, taking into account also potential variability present within the individual observations, and free from 
instrumental variations that are not accounted for by the \emph{dmextract} tool\footnote{\href{https://cxc.cfa.harvard.edu/ciao/threads/variable/}{https://cxc.cfa.harvard.edu/ciao/threads/variable/}} typically used to extract light curves.
Pre-EWOCS observations were excluded due to the significant time gap between them and the EWOCS observations, during which the ACIS detector's sensitivity to soft X-rays declined significantly.

To search for periodic signals in the lightcurve, we constructed a Lomb-Scargle (LS) periodogram which is a parametric method designed to detect and characterize periodic signals in unevenly spaced time-series data by fitting sinusoidal models at a range of trial frequencies and evaluating their goodness of fit  \citep{zechmeister09,vanderplas12,vanderplas15}. 
More specifically we utilised the \textit{astropy} \texttt{LombScargle} class\footnote{\href{https://docs.astropy.org/en/stable/api/astropy.timeseries.LombScargle.html\#id1}{https://docs.astropy.org/en/stable/api/astropy.timeseries.LombScargle.html}} and used the \texttt{.autopower()} function with \texttt{nyquist\_factor=10} and \texttt{samples\_per\_peak=1000}.
The 1$\sigma$ uncertainty of the strongest peak is estimated from the full width at half maximum (FWHM) of the peak. The false alarm probability is calculated using the built-in bootstrap method of the \textit{astropy} LS class, with a hundred thousand iterations.
We also computed the window function (WF) by calculating an LS periodogram of a flat signal (constant value of 1) with the same time sampling as our light curve, using \texttt{fit\_mean=False} and \texttt{center\_data=False}\footnote{As suggested under “additional arguments” at \href{https://docs.astropy.org/en/stable/timeseries/lombscargle.html}{https://docs.astropy.org/en/stable/timeseries/lombscargle.html}}. 
For comparison, we normalized the WF so that its highest peak aligns with the main peak detected in our data. 
In this way, the other peaks in the WF reveal the alias structure introduced by the uneven sampling, helping us to distinguish genuine periodicities from spurious ones in the LS periodogram.

To further validate our analysis using a method entirely independent of the Lomb-Scargle approach, we applied the String Length Method \citep{dworetsky83}. In particular we used the Python implementation provided in the \texttt{PyAstronomy} package\footnote{\href{https://pyastronomy.readthedocs.io/en/latest/pyTimingDoc/stringlength.html}{https://pyastronomy.readthedocs.io/en/latest/pyTimingDoc/stringlength.html}}.
The String line method is a non-parametric approach that searches for the period that minimizes the total ‘string length’—the sum of distances between consecutive points when the lightcurve is folded over trial periods. A clear minimum in string length indicates a likely true period.
Moreover for a more conservative approach on the uncertainty of a period we used the Celerite-powered Gaussian Process (GP) framework implemented in the \textit{exoplanet} toolkit\footnote{\href{https://docs.exoplanet.codes/en/v0.1.3/}{https://docs.exoplanet.codes/en/v0.1.3/}} which is specifically optimized for scalable modelling of time-series data. Before fitting, the light curve was standard-normalized to zero mean. The Gaussian Process was defined as a combination of a kernel accounting for  the periodic changes and another term describing the excess white noise. The periodic kernel was initialized using the period corresponding to the most prominent peak in the LS periodogram \footnote{for examples: \href{https://gallery.exoplanet.codes/tutorials/stellar-variability/}{ https://gallery.exoplanet.codes/tutorials/stellar-variability/}}.
The results on the variability and periodicity investigation are presented in Section\,\ref{periodresults}.

\subsection{Deriving X-ray colours}

We determine the X-ray colours and their associated $1\sigma$ uncertainties, utilizing the \texttt{BEHR} tool \citep[Bayesian Estimation of Hardness Ratios;][]{park06}. This method provides statistically robust estimates of X-ray colours and their confidence limits, even when one or both bands contain very few counts as it evaluates the full posterior probability distribution of the colour, accounting for both source and background counts under Poisson statistics. We opted for colours rather than classical hardness ratios because they yield more symmetric posterior distributions in the low count regime \citep[see Fig. 8 in][]{park06}.
We define the X-ray colours as follows: $C_1 = \log_{10}(\mathrm{S}/\mathrm{M})$, $C_2=\log_{10}(\mathrm{M}/\mathrm{H})$, and $C_3 = \log_{10}(\mathrm{S}/\mathrm{H})$, where $\mathrm{S}$, $\mathrm{M}$, and $\mathrm{H}$ represent the net counts in the soft (0.5--2.0\,keV), medium (2.0--4.0\,keV), and hard (4.0--8.0\,keV) bands, respectively. The X-ray colours of Wd1-9 are presented in Section\,\ref{coloursresults}.

\subsection{Spectral analysis}

For the spectral analysis, all available observations (including EWOCS and pre-EWOCS data) were jointly analysed. Given that the spectral characteristics of Wd1-9 can change over time, especially if it is a binary, the resulting fits represent average measurements. Detailed information on photon extraction and background region selection can be found in \citet{guarcello24}. 
Given the high source counts, we used \chisq{} statistics after grouping the data with a minimum of 15 counts per bin to ensure reliable fitting.

We modelled the combined spectrum in XSPEC \citep[v.12.13.0c;][]{arnaud96} using two approaches: Model A (\texttt{tbabs$\times$phabs$\times \sum$\texttt(vapec)}) and Model B (\texttt{tbabs$\times$phabs$\times \sum$\texttt(vpshock)}). Both models incorporate two absorption components: \texttt{tbabs}, which accounts for ISM absorption towards Wd1, fixed at $2.05\times10^{22}$\,atoms\,cm$^{-2}$ \citep[HEASARC \nh{} tool;][]{nhtool}, and \texttt{phabs}, which represents the absorbing column potentially local to Wd1 or within the stellar wind where elemental abundances may differ from those assumed in the ISM-focused \texttt{tbabs} model.
In each model, \texttt{vapec} and \texttt{vpshock} describe different types of emission spectra: \texttt{vapec} assumes collisionally ionised plasma in equilibrium, while \texttt{vpshock} models a constant-temperature, plane-parallel shock that can accommodate non-equilibrium conditions. In both models, varying the elemental abundances was an option to improve the fit to the data. We utilised the abundance table from \citet{wilms00}, as it is the most updated option for the \texttt{tbabs} model. The spectral analysis results are reported in Section\,\ref{spectralresults}. All uncertainties are quoted at the 1$\sigma$ confidence level.

\section{Results}\label{analysisandresults}

\begin{figure}[!h]
        \includegraphics[width=1.0\columnwidth]{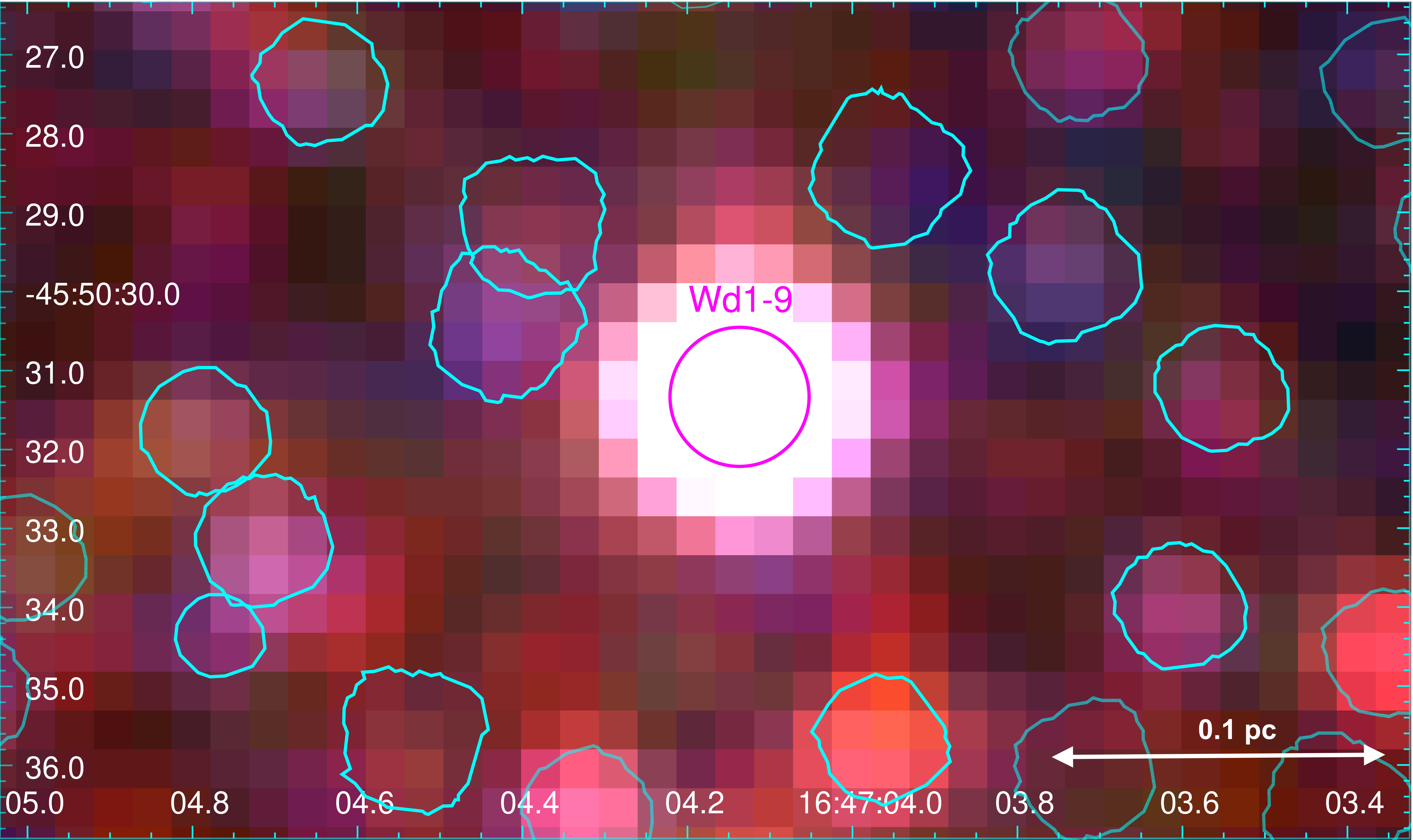}
        \caption{The combined smoothed \chandra{} colour image of Wd1-9. X-ray photons are colour-coded based on their energy range: soft (0.5–2.0 keV) appear in red, medium (2.0–4.0 keV) in green, and hard (4.0–8.0 keV) in blue. We show with a purple circle the extraction region of Wd1-9, while nearby X-ray detected sources are marked with cyan.
        } 
 		\label{fig.colorimagew9}
 \end{figure}

Wd1-9 (CXO J164704.1-455031; RA: 16:47:04.13; Dec: $-$45:50:31.3) is the third brightest X-ray source in Wd1 (Fig.\,\ref{fig.colorimagew9}), following the magnetar CXO J164710.20-455217 and the the Wolf-Rayet (WR) star A \citep{guarcello24,konna24}. It has $7975\pm90$ net counts in the 0.5--8\,keV band among a total of 5963 validated X-ray sources  \citep{guarcello24}.
The bright X-ray nature of Wd1-9, combined with the very deep \chandra{} exposure, allows for the first time a detailed spectrum, colour, and timing analysis, which we describe in the following Sections. 

\subsection{Variability and periodicity detection}\label{periodresults}

\begin{figure}[!h]
   \includegraphics[width=1.0\columnwidth]{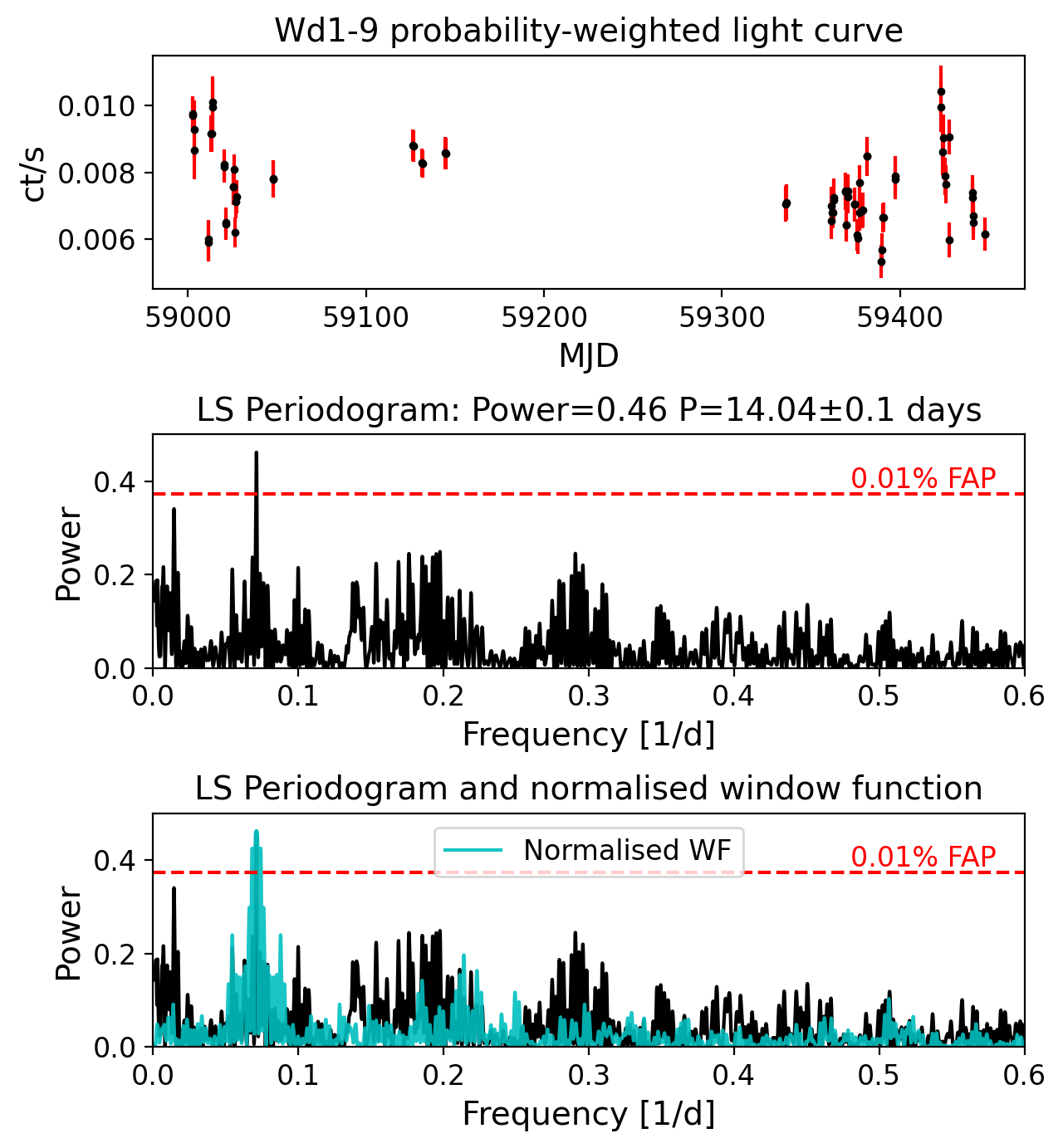}
        \caption{
        EWOCS light curve and periodogram for Wd1-9. Top panel:  Broad-band probability-weighted light curve generated using the \emph{glvary} tool. Middle panel: LS periodogram with a red dashed line indicating the 0.01\% false alarm probability (FAP). Bottom panel: Same as the middle panel, with the normalised window function superimposed on the periodogram.} 
 		\label{fig.lcbroadw9}
 \end{figure}

We constructed a long-term light curve using the probability-weighted light curve with optimal time binning from the \textit{glvary} tool\footnote{Section 3.3 of the second EWOCS paper provides a detailed description of the \textit{glvary} analysis \citep{konna24}.} (top panel of Fig.\,\ref{fig.lcbroadw9}).
To assess long-term variability, we initially performed a chi-square test which resulted in the source being variable at a 99\% confidence level. We then constructed the LS periodogram which we present in the middle panel of Fig.\,\ref{fig.lcbroadw9}. We show with a horizontal dashed line the 0.01\% false alarm probability (FAP) level. The LS periodogram exhibits a significant period at $14.04\pm0.1$ days. This is the first time a period has been detected for this system. We also overplot the WF (cyan colour; bottom panel in Fig.\,\ref{fig.lcbroadw9}), normalised at the main periodicity. 

\begin{figure}[!htbp]
  \includegraphics[width=0.9\columnwidth]{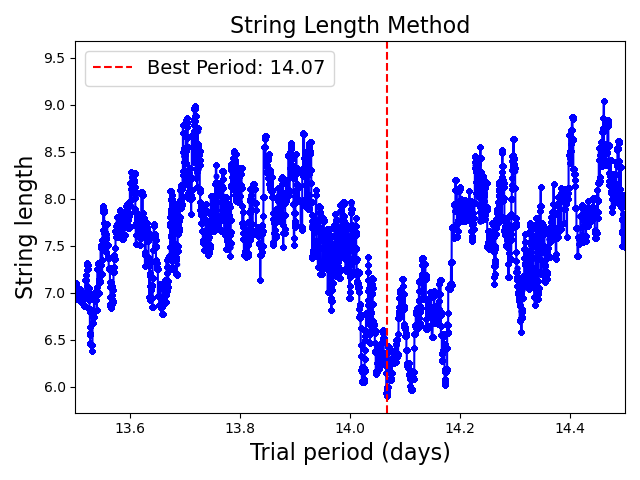}
 \caption{String length method applied to determine and verify the best period. The blue points represent the string length values for different trial periods, while the red dashed line indicates the best period found at 14.07 days. This plot focuses on the trial period range of 13.5--14.5 days, providing a detailed view of the minimum.}
 		\label{fig.string}
 \end{figure}

A peak observed at approximately 79 days shows a much weaker signal in the WF but falls below the FAP threshold. To explore if this could be a real periodicity, we investigated potential secondary periodicities after removing the primary signal, but no significant peaks were identified. 
Moreover, using the String Length method, performed with 5 million samples over a trial period range of 10 to 40 days, consistently recovered the same period of 14.07 days. The smallest string lengths were found within the range of 14.02 to 14.18 days. A zoomed-in view is provided in Fig.\,\ref{fig.string}.

\begin{figure}[!htbp]
  \includegraphics[width=0.9\columnwidth]{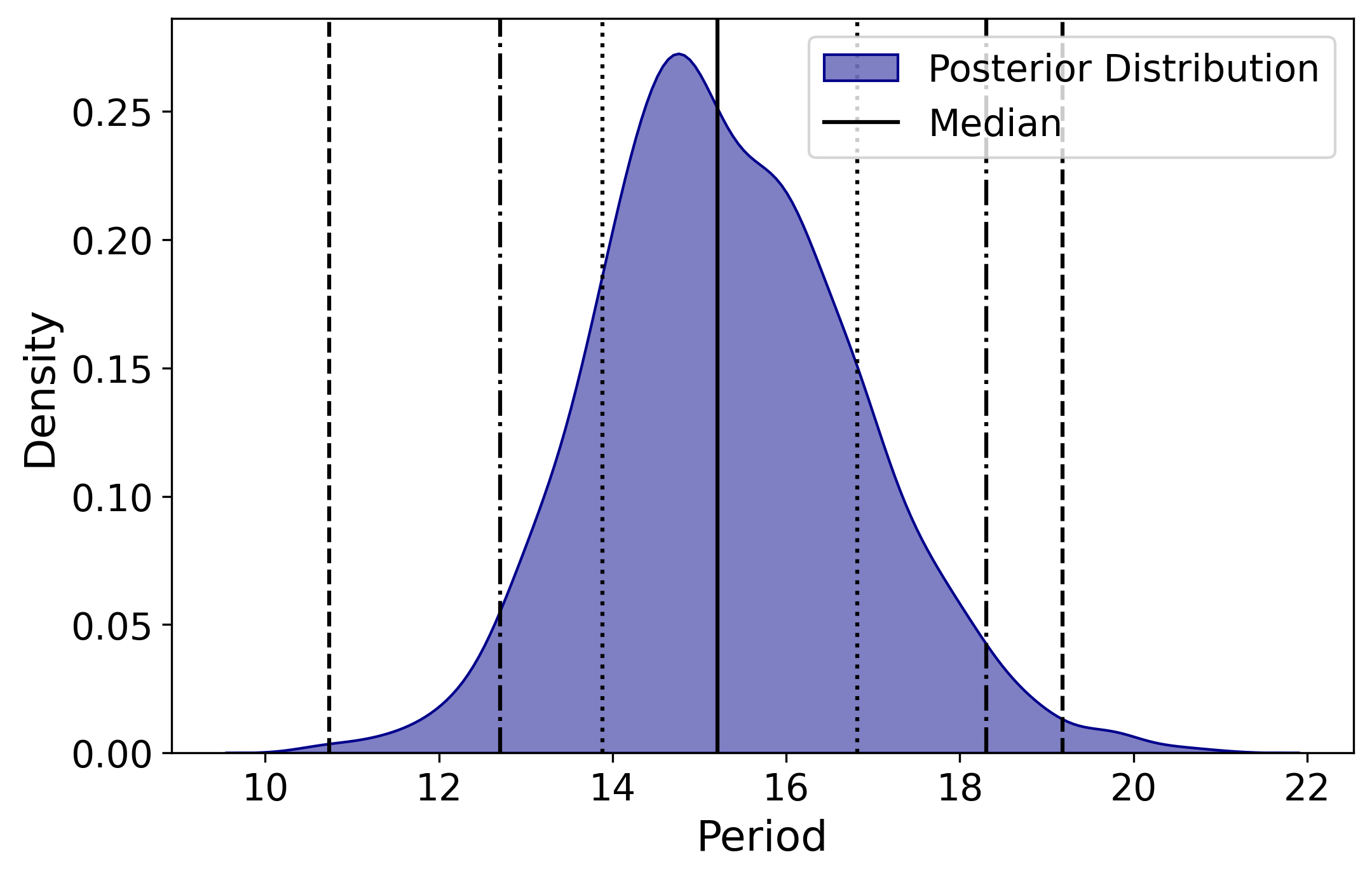}
 \caption{Posterior distribution for the detected period using the GP method. The median value (i.e. the central period) is shown with the solid line, while the 16--84th, 3--97th, and 1--99th percentiles are shown with the dotted, dash-dotted, and dashed lines, respectively.}
 		\label{fig.gp}
 \end{figure}

To adopt a more conservative approach for estimating the period uncertainty, we utilized the Celerite-powered GP described in Section \ref{varmethods}. The posterior distribution for the period is presented in Fig.\,{\ref{fig.gp}} and provides the median value (15.2\,d; solid line) which is the central period estimate, the 16--84th (13.8--16.8\,d; dotted lines), the 3--97th (12.4--18.3\,d; dash-dotted lines), and the  1--99th (10.7--19.2\,d; dashed lines) percentiles to reflect its asymmetric uncertainty. The posterior distribution is slightly skewed toward longer periods, which could be a result of limited phase coverage, noise, or irregular sampling, all of which can allow a broader range of plausible periods. We note that the period derived from the GP modelling is consistent, within the 16th to 84th percentile range, with the values obtained from both the LS and string length methods. For consistency, we adopt the LS-derived period (14.04 days) as the reference value throughout the remainder of the manuscript.

\begin{figure}[!htbp]
  \includegraphics[width=0.9\columnwidth]{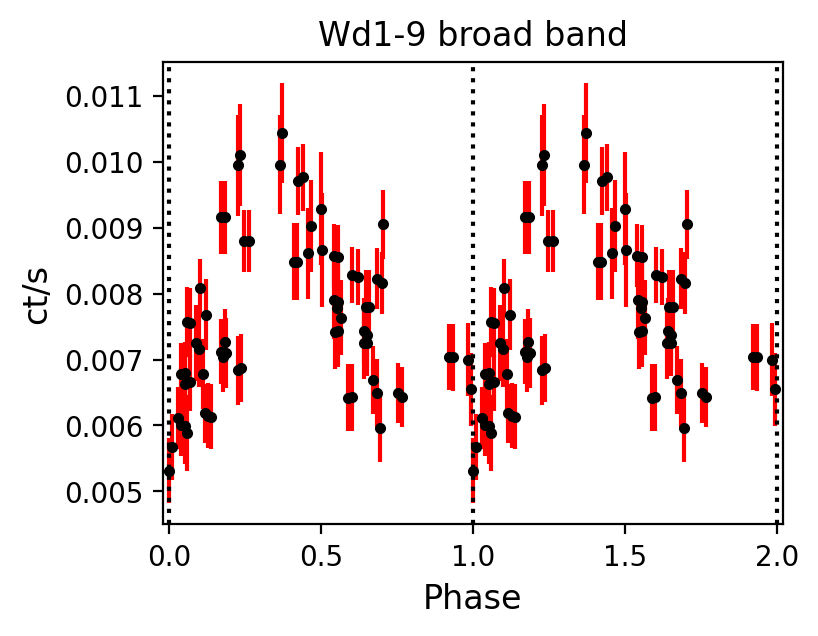}
 \caption{Broad-band unbinned phase-folded probability-weighted light curve of Wd1-9.}
 		\label{fig.phasedlc}
 \end{figure}

To examine whether the detected periodicity is connected to some specific energy band, we produced soft (0.5--2\,keV), and hard (2--8\,keV) energy band light curves and LS periodograms. Both resulted in the detection of the exact same period of $\sim$14 days. 
Fig.\,\ref{fig.phasedlc} presents the broad-band, unbinned, phase-folded, probability-weighted   light curve for Wd1-9, based on the most significant period identified by the LS method. Phase 0 is set at the minimum flux in the broad band.

\subsection{Properties of X-ray colours}\label{coloursresults}

\begin{figure}[!h]
 	\centering	
        \includegraphics[width=1.0\columnwidth]{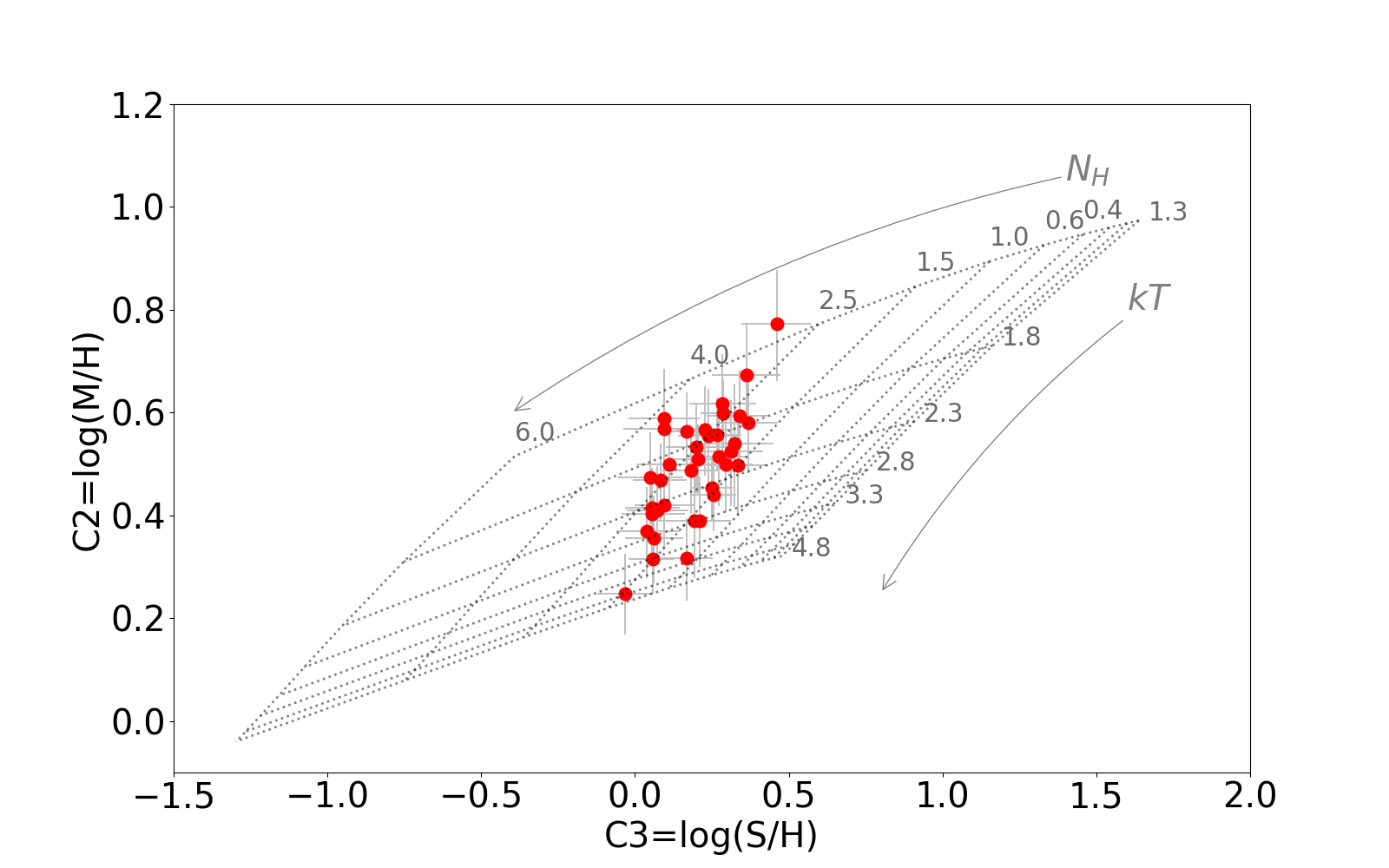}
        \caption{X-ray colour-colour diagram for Wd1-9. $\mathrm{S}$, $\mathrm{M}$, and $\mathrm{H}$ represent the net counts in the soft (0.5--2.0\,keV), medium (2.0--4.0\,keV), and hard (4.0--8.0\,keV) bands, respectively. The 36 EWOCS observations are shown with red circles along with their 1$\sigma$ error bars. The grey-dotted grid and the \nh{} and $kT$ values correspond to the results of simulated absorbed thermal spectra (model used:\texttt{tbabs}$\times$\texttt{apec}). 
        } 
 		\label{fig.colorw9}
 \end{figure}

We present in Fig.\,\ref{fig.colorw9} the $C_2$ versus $C_3$ colours of Wd1-9 for the 36 EWOCS observations, overlaid on a simulated grid of absorbed thermal spectra (grey lines) showing the direction of increasing plasma temperature ($kT$) and increasing absorption (\nh). 
All observations fall on top of the grid and are predominately hard in nature.
The estimated plasma temperatures range from 1.3 to 4.8\,keV,  and given the associated uncertainties, it is likely that different values trace true physical changes in the temperature of the emitting plasma.
The absorption has an average value of $\sim$2.5$\times10^{22}$ atoms\cmsq within uncertainties.

\subsection{Spectral results}\label{spectralresults}

\begin{figure}[htbp]
 	\centering
        \includegraphics[width=1\columnwidth]{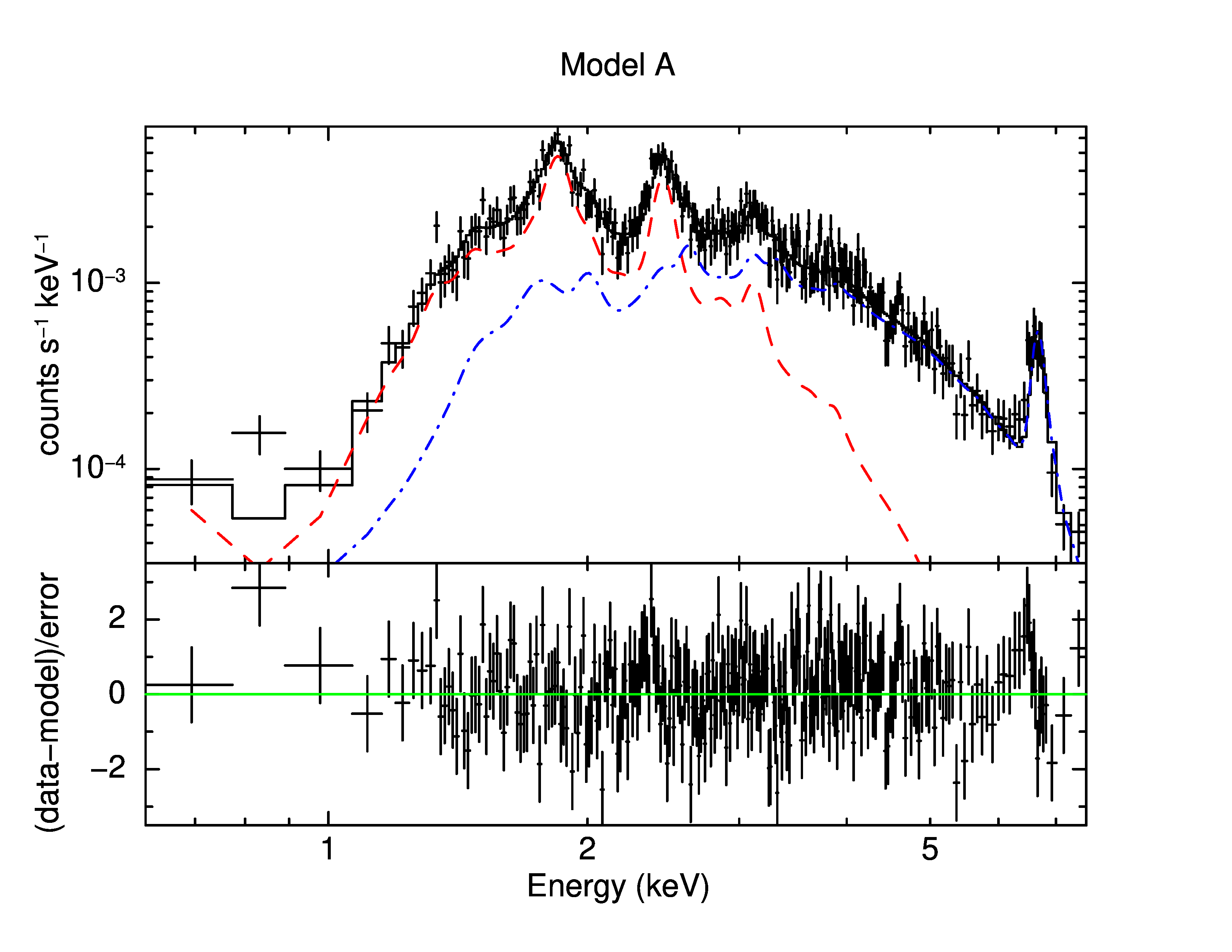}  
          \includegraphics[width=1\columnwidth]{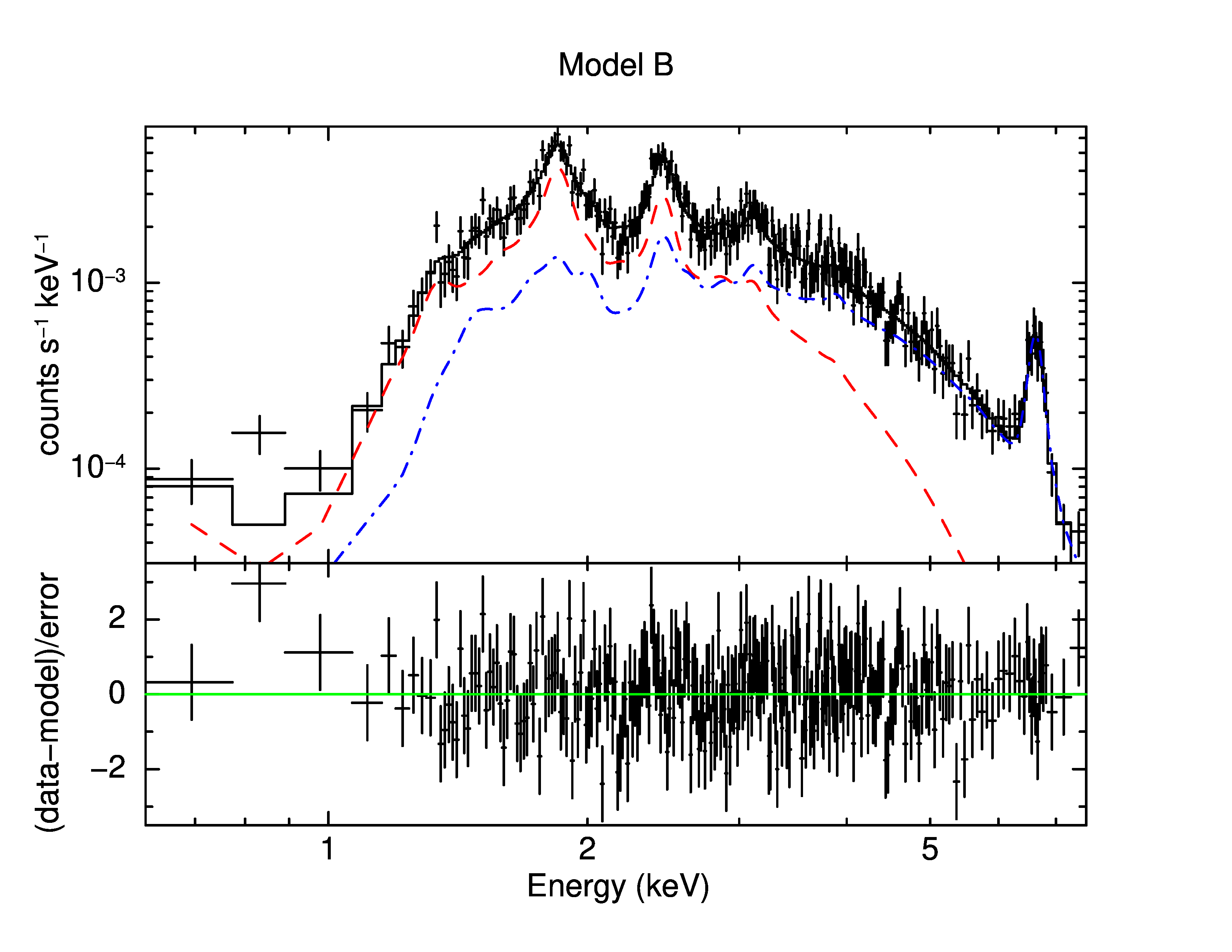}
           \caption{Spectral modelling of Wd1-9 using Model A (top) and Model B (bottom). The contributions of the soft and hard thermal components are shown with red dashed and blue dash-dotted lines, respectively. The errors correspond to the 1$\sigma$ confidence and the residuals are reported in the bottom panel in terms of sigma.}
 		\label{fig.combinedspectra}
 \end{figure}

We present the best-fit models and spectral fitting results for Models A and B in Fig.\,\ref{fig.combinedspectra} and Table\,\ref{tab.spectralfit_solar}. The spectrum exhibits strong emission lines of  Si~XIII ($\sim$1.8\,keV),  S~XV ($\sim$2.5\,keV), Ar~XVII ($\sim$3.1\,keV), and  Fe~XXV  ($\sim$6.7\,keV) which is observed for the first time in the EWOCS data. The emission lines are fitted well by two thermal components, one softer ($\sim$0.7--0.9\,keV), and one harder ($\sim$3.0\,keV). The fit of the \texttt{vpshock} model was constrained to adopt the same ionisation parameter ($\tau$) for both thermal components. In the initial fitting trials, significant residuals were present at energies around certain emission lines, and allowing the corresponding abundances to vary (always tied between the two thermal components) led to a substantial improvement in the fit. Namely, in the thermal components of Model A, the abundances of S and Ar were allowed to vary freely, whereas in Model B only the abundance of Fe was treated as a free parameter (see notes in Table\,\ref{tab.spectralfit_solar}). 
Both models fit the spectra well, including the emission lines, with \chisqr close to unity.
Model B exhibits smaller residuals around the Fe~XXV line, while both models show a soft data excess at energies 0.7--0.9\,keV. 
To quantitatively compare the two spectral models, we computed the Akaike Information Criterion (AIC; \citep{akaike74}) and the Bayesian Information Criterion (BIC; \citep{schwartz78}). Model A, with 8 free parameters, yields AIC = 289.2 and BIC = 317.7, while Model B, with 7 free parameters, gives AIC = 265.8 and BIC = 290.7. Both criteria favour Model B, indicating it provides a significantly better fit to the data.
The soft excess persists even when the archival data are excluded from the combined spectrum, minimizing the impact of potential calibration differences due to the time gap between the EWOCS and archival observations.

\begin{table*}[htbp]
	\centering
		\caption{X-ray spectral fitting results for the combined spectrum}
            \begin{tabular}{@{}llllllcc@{}}
			\toprule		
		Model &  $N\rm{_{H}^{local}}$ & $kT$ & norm & $\tau$  &$\chi^2_{\nu}$ ($\chi^2$/dof)  & $F_X$[$F\mathrm{_X^{ISMcor}}$] &  $L_X$[$L\mathrm{_X^{ISMcor}}$] \\[2pt]
&  $10^{22}$\,\cmsq  &  keV  & cm$^{-5}$ & $10^{10}$\,s\,cm$^{-3}$ &&  10$^{-13}$\funit & 10$^{32}$\ergs\\
\midrule
A\tablefootmark{a} & 1.46$_{-0.18}^{+0.23}$ &0.67$\pm0.09$ &(0.81$_{-0.19}^{+0.37}$)e$-$3  &&1.08 (273.22/253)& 1.45$_{-0.05}^{+0.01}$[2.8] &3.11$_{-0.11}^{+0.02}$[5.99] \\ [3pt]
 &  &  2.97$_{-0.35}^{+0.51}$ &(1.75$_{-0.32}^{+0.35}$)e$-$4  &&&& \\ [3pt]
B\tablefootmark{b} & 4.26$_{-0.27}^{+0.31}$ &0.85$_{-0.16}^{+0.19}$ &(1.07$_{-0.21}^{+0.44}$)e$-$3& 17.82$_{-4.79}^{+7.96}$  &1.0 (251.78/253)&1.45$_{-0.22}^{+0.04}$[2.06] &3.11$_{-0.48}^{+0.09}$[4.41] \\ [3pt]
 &  & 3.19$_{-0.74}^{+3.01}$ &(1.71$_{-0.96}^{+1.0}$)e$-$4  &&&& \\ [3pt]
\bottomrule
    \end{tabular}
    \tablefoot{Model A corresponds to \textit{tbabs$\times$ phabs$\times \sum$vapec}, and  Model B corresponds to \textit{tbabs$\times$ phabs$\times \sum$vpshock}. The plasma temperature is expressed in keV, with a normalization parameter defined as $\frac{10^{-14}}{4\pi D^2}\int n_e n_H , dV$, where $n_e$ and $n_H$ denote the electron and hydrogen densities, integrated across the emitting volume $V$. Here, $D$ represents the distance to the source in centimetres \citep{smith01}. $\tau$ is the upper limit on the ionization time scale. We show the absorbed 0.5--8\,keV flux, the corresponding luminosity, and in brackets the ISM-corrected values. All uncertainties are quoted at the 1$\sigma$ confidence level. Abundances of specific elements that were varied are as follows: \tablefoottext{a}{S=2.03$\pm0.30$, Ar=2.68$\pm0.9$}\tablefoottext{b}{Fe=3.52$_{-0.96}^{+1.25}$}  
    \label{tab.spectralfit_solar}}
\end{table*}

\subsection{Phase-resolved spectral properties}

\begin{figure}[htbp]
\includegraphics[width=0.9\columnwidth]{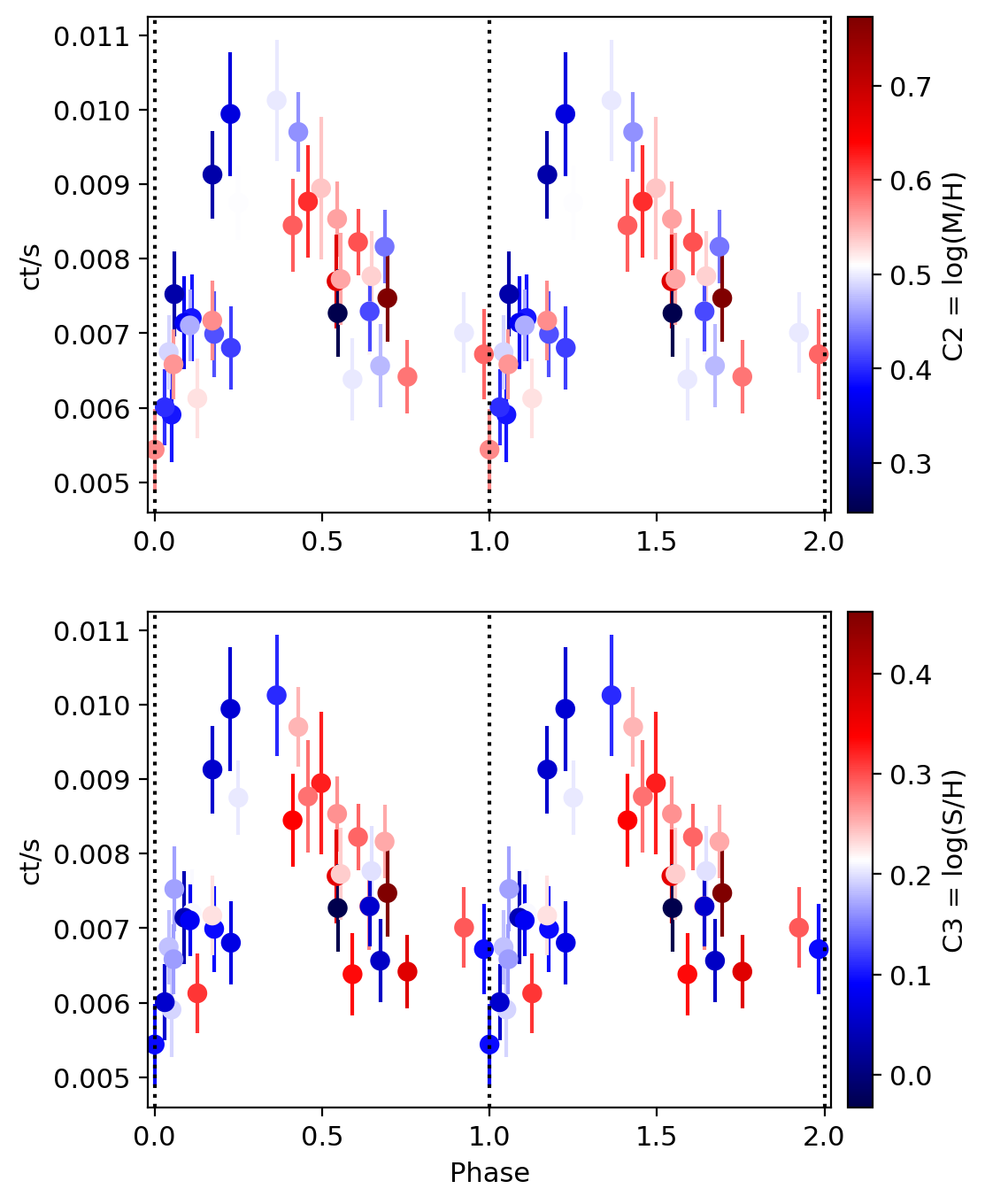} 
 \caption{Phase-folded X-ray colour-coded broad-band light curves for the X-ray colours $C_2$ (top panel) and $C_3$ (bottom panel).}
 	\label{fig.phasedlccolor}
 \end{figure}

To explore potential correlations between the phase and X-ray emission, we plotted phase-folded light curves, colour-coded according to the X-ray colours. For the $C_1$ colour we did not observe any specific trend. 
However, the X-ray colour-coded phase-folded light curves for colours $C_2$ (top panel in Fig.\,\ref{fig.phasedlccolor}) and $C_3$ (bottom panel Fig.\,\ref{fig.phasedlccolor}) reveal that harder colours generally exhibit their peak fluxes at earlier phases compared to the softer colours. This behaviour is observed both between the soft($<$2\,keV) and hard ($>$4\,keV) bands, as well as between the medium (2--4\,keV) and hard ($>$4\,keV) bands.  Therefore, this may indicate some physical or observed change in the system at different phases for energies below and above 4\,keV.

\begin{figure*}[htbp]
\includegraphics[width=0.7\columnwidth]{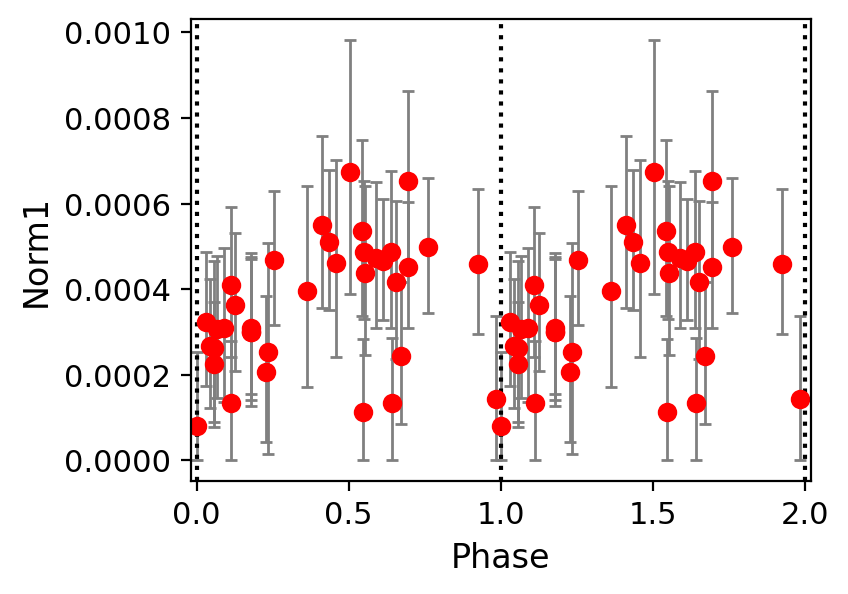}  \includegraphics[width=0.7\columnwidth]{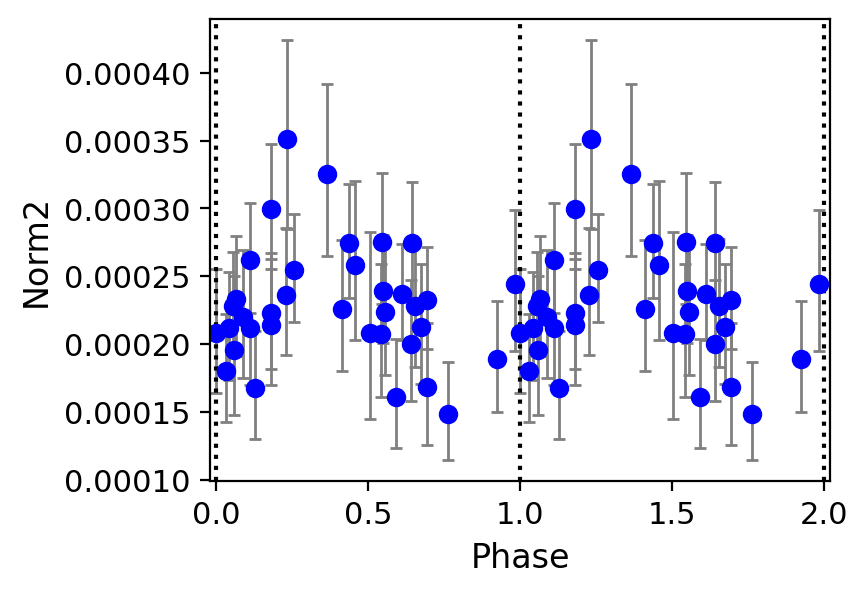}    \includegraphics[width=0.6\columnwidth]{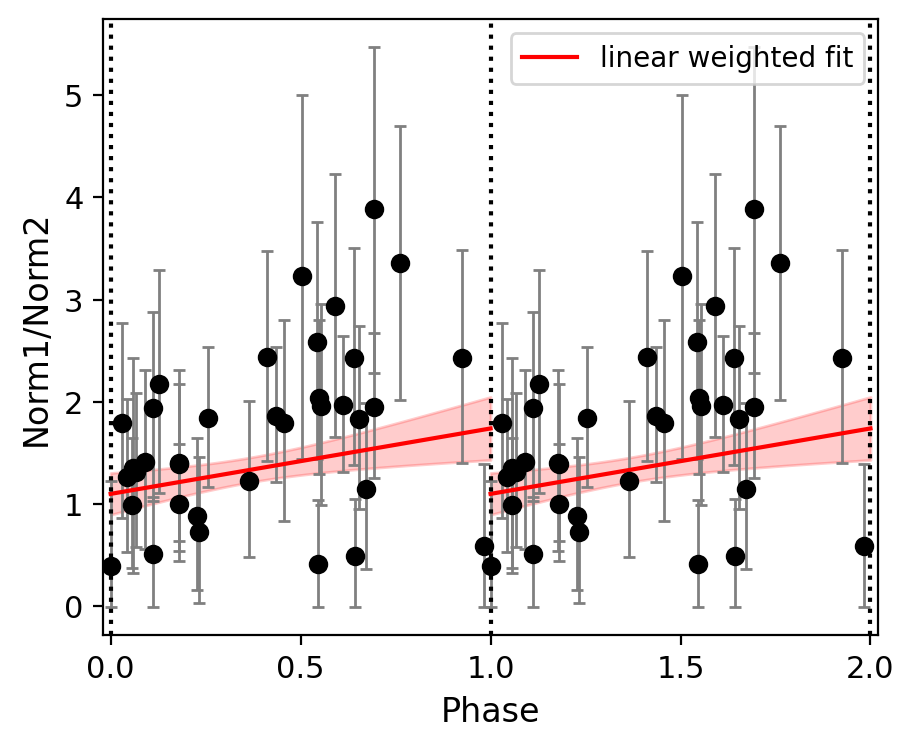}
 \caption{Phase-folded light curves of the normalisations of the thermal components and of their ratio. Left: the variation over phase of the  normalisation (norm1) of the soft thermal component ($kT$=0.67\,keV ). Middle: Same as left panel but for the hard thermal component (norm2; $kT$=2.97\,keV) Right: The variation over phase of the  normalisation ratio (norm1/norm2) and its weighted linear fit shown with the red solid line}. 		\label{fig.phasedlcnorm}
 \end{figure*}

X-ray colours are effective indicators of a star's spectral behaviour. However, considering that the Wd1-9 spectrum can be adequately modelled by two thermal components (Table\,\ref{tab.spectralfit_solar}), both components contribute at energies below and above 4\,keV, which corresponds to the region where we observe a distinct trend in the phase-folded light curves (Fig.\,\ref{fig.phasedlccolor}). 
To disentangle the contributions of the soft and hard spectral components, we independently fitted the spectra of each of the 36 EWOCS observations using Model A, since it is the simplest model of the two, keeping all components fixed to their best-fit values (Table\,\ref{tab.spectralfit_solar}) except for the normalizations of the thermal components, which were allowed to vary.
This is a simplifying approximation, as we acknowledge that variations, particularly in a binary system, may be expected in the absorption and temperatures of the thermal plasmas. However, given the low statistics of the individual spectra, which do not allow for a more detailed fit, this provides a reasonable solution. In Fig.\,\ref{fig.phasedlcnorm} we present the phase-dependent results of the normalizations for the two thermal components and their ratio along with the weighted linear fit. 
The normalization of the soft thermal component (right panel in Fig.\,\ref{fig.phasedlcnorm}) reaches its peak at a later phase than the hard thermal component (middle panel in Fig.\,\ref{fig.phasedlcnorm}). This pattern aligns with the phase-folded X-ray colour-coded light curves, which show similar phase-dependent behaviour in Fig.\,\ref{fig.phasedlccolor}. Moreover, a linear weighted fit reveals that the ratio between the two normalizations increases steadily with phase as $Norm1/Norm2 = 0.6406 \times phase + 1.0974$ making clear that the soft thermal component dominates at later phases.
One difference when accounting for each spectral component instead of energy bands is that the emission of the soft thermal spectral component is higher than that of the hard spectral component for most observations and by as much as four times for phases between 0.7 and 1.0 (right panel in Fig.\,\ref{fig.phasedlcnorm}). 
This is expected and reflects the methodological differences, as the emission from the soft spectral component can extend beyond 2\,keV which is the upper limit of the soft colour band.
We also examined the phase-binned spectra of the source for possible variations and we present the results of this analysis in the appendix.

\subsection{Spectral phase map of Wd1-9}

\begin{figure*}[htbp]
\includegraphics[width=1.9\columnwidth]{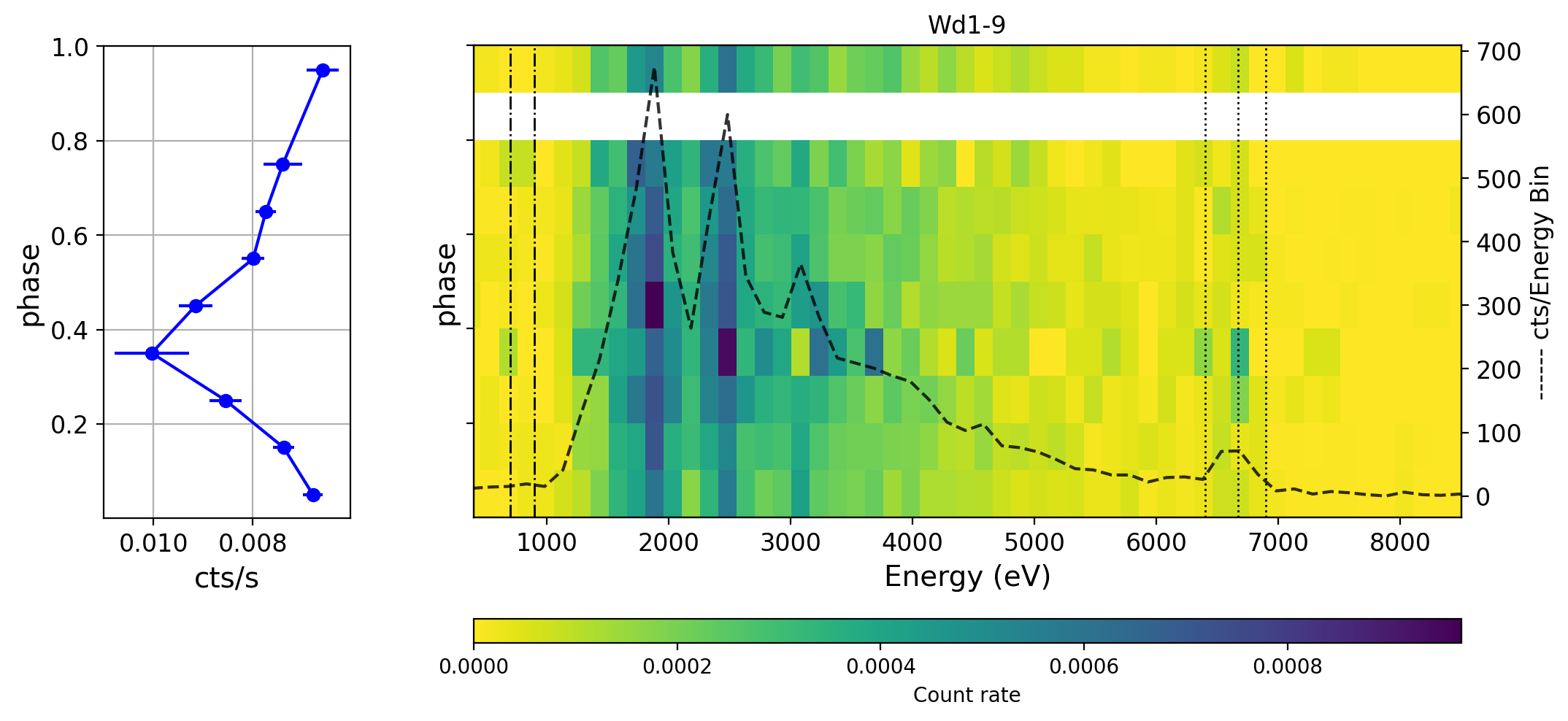}
  \caption{  
Spectral phase map of Wd1-9 using all EWOCS observations. 
Left Panel: The phase-folded light curve of Wd1-9, highlighting the contribution of broad-band events in the various phase bins.
Right Panel: The phase-folded combined event file, binned over phase and energy range and normalised by the corresponding exposure time. Different colours represent varying count rates, as indicated by the colour bar below. 
The white stripe represents a phase bin with no available observations.
The dashed line illustrates the summed counts in each energy bin (exact values are shown at the right y-axis), providing a general representation of the spectrum and the dominant emission lines. The dotted lines on the right mark the peak emission from Fe~K$\alpha$, Fe XXV, and Fe XXVI lines. On the left, dash-dotted lines indicate the energy range where excess soft emission is observed. }
 \label{fig.dynamicspec}  
\end{figure*}

To gain a clearer understanding of how individual emission lines contribute to the overall spectrum of Wd1-9, we analysed the evolution of the event files across different phases \citep[see, for example, figure 3 in][]{ness22}.
In Fig.\,\ref{fig.dynamicspec}, the main panel on the right displays the phase evolution of the combined EWOCS event file which was constructed from broad-band event files across 36 observations, binned into 0.1 phase intervals and grouped into energy bins of 150 eV, which is comparable to, and slightly oversamples, particularly at lower energies ($\sim$1.49\,keV), the on-axis energy resolution of ACIS-I\footnote{The \chandra{} Proposer’s Observatory Guide, v. 27.0}.

 Different colours indicate the count rates, as shown on the accompanying colour bar. The dashed line represents the total counts summed across each energy bin, with corresponding values displayed on the right y-axis. This line provides an overview of the spectrum and highlights the prominent emission lines. Dotted lines on the right denote the peak emission from Fe~K$\alpha$, Fe XXV, and Fe XXVI, while the dash-dotted lines on the left mark the energy region associated with excess soft emission.
The left panel presents the phase-folded light curve of Wd1-9, emphasizing the contribution from broad-band events in the corresponding bins.

\begin{figure*}[!h]
\includegraphics[width=1.9\columnwidth]{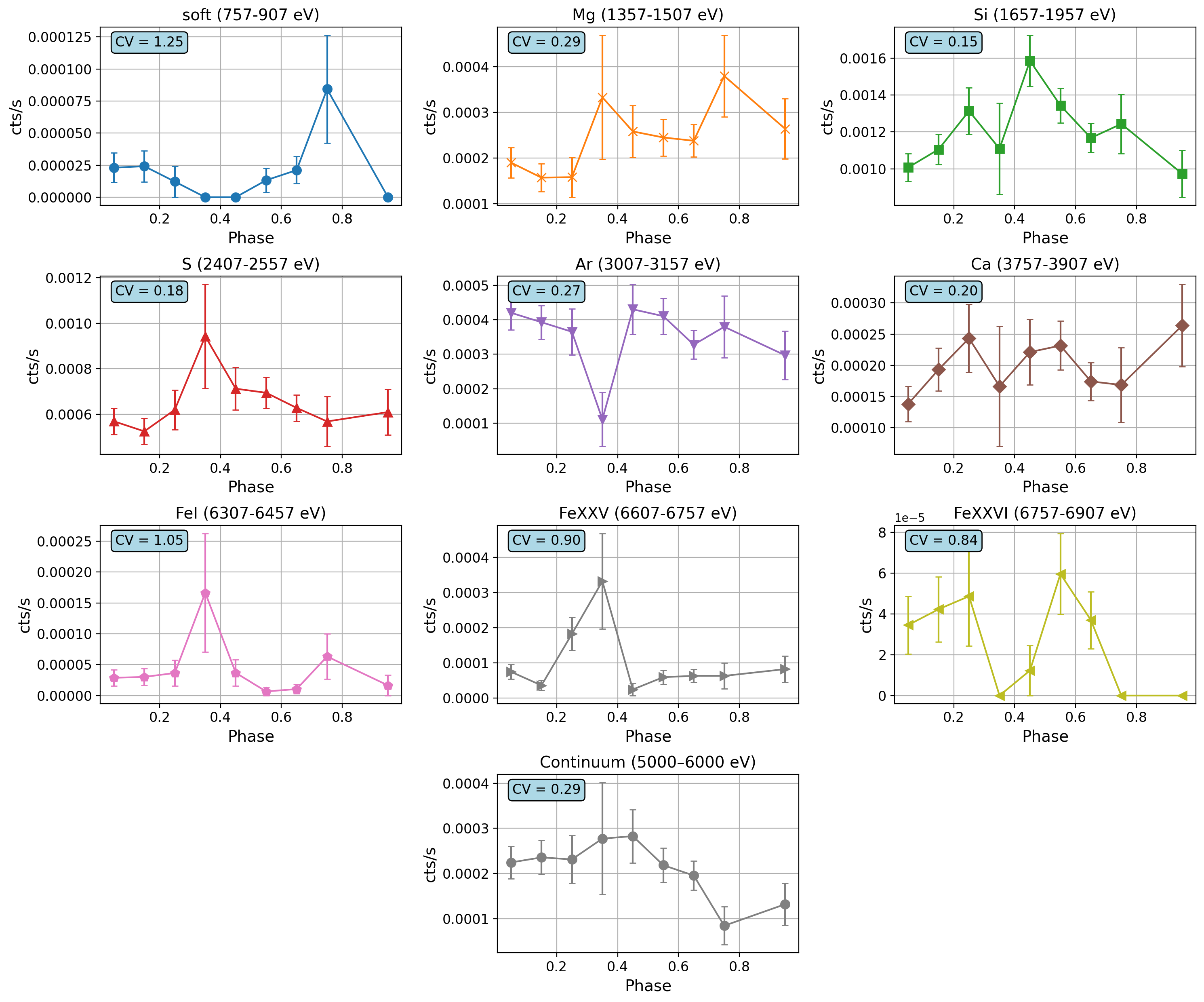}
\caption{Phase-dependent variation of the total (line plus continuum) count rate in energy bands centred on selected emission lines. The coefficient of variation (CV) is defined as the ratio of the standard deviation to the mean count rate, serving as a measure of emission variability across phase in each energy range. The last panel shows the light curve extracted from a line-free energy range (5.0--6.0\,keV), representative of the underlying continuum.}
\label{fig.variationoflines}
\end{figure*}

In Fig.\,\ref{fig.variationoflines}, we illustrate the variation in the count rates of the selected emission lines as a function of phase. We used the spectral phase map and extracted the count-rates for the energy ranges including the peak emission of each line. The adopted energy ranges are shown in the title of each panel in Fig.\,\ref{fig.variationoflines}.
We also show on the top left label of each panel the coefficient of variability (CV), which quantifies the fluctuations in the data relative to its mean. CV is defined as as the ratio of the standard deviation to the mean count rate (CV = $\sigma / \mathrm{mean}$).
The variability of Mg, Si, S, Ar, and Ca emission is relatively low, with the standard deviation being less than 30\% of the mean count rate. Overall,  in our analysis, we consider $\mathrm{CV} < 0.3$ as indicating \textit{low variability}, $0.3 \leq \mathrm{CV} < 0.6$ as \textit{moderate variability}, and $\mathrm{CV} \geq 0.6$ as \textit{significant variability}.
Most of the strong emission lines exhibit variations around phase 0.35, where the overall flux reaches its peak. Specifically, sulphur (S) and iron (Fe~K$\alpha$ and Fe XXV) show an increase in flux at this phase, while argon (Ar) and iron Fe XXVI exhibit a decrease. The Fe~K$\alpha$ line shows also a small increase between phases 0.5--0.8 which also present in bin 3  of the phased  spectra (Fig.\,\ref{fig.phasedspectrazoom}) in the appendix. In contrast, magnesium (Mg) and calcium (Ca) do not display any pronounced variations and silicon (Si) shows a peak at phase=0.5 which coincides with the peak emission of the soft spectral component (Fig.\,\ref{fig.phasedlcnorm}). Additionally, we analysed the soft emission between 750 eV and 900 eV, where an excess is observed in the combined spectrum (see Fig.\,\ref{fig.combinedspectra}). The soft emission shows a slight decrease in count rate at phase 0.4 and a more pronounced increase at phase 0.75 in agreement with the increased soft excess observed in bin 1 of the phased spectra (Fig.\,\ref{fig.phasedspectra} in the appendix). 
We should note here that the count rates shown for each emission line represent the total emission (i.e., the combined contribution from both line and continuum) within the selected energy intervals. To assess the relative role of the continuum, we also extracted a light curve from a line-free region between 5.0--6.0\,keV. This band shows no significant variability up to phase 0.65, where the strongest modulations are observed in bands containing prominent emission lines. This suggests that the observed variability in those bands is primarily driven by changes in line emission rather than variations in the underlying continuum.

\section{Discussion}\label{discussion}

\subsection{Comparison with previous X-ray works}

\citet{skinner06} using the first $\sim$60\,ks \textit{Chandra} observations of Wd1 detected Wd1-9 with $\sim$330 counts and reported an unabsorbed luminosity which, scaled for our adopted distance, is an order of magnitude higher ($L_X = 4.48 \times 10^{33}$ \ergs) than our average measurements in Table\,\ref{tab.spectralfit_solar}. This difference can be attributed to \citet{skinner06} using PIMMs with a conversion factor based on a softer spectrum (kT = 1.0\,keV) than the one actually observed ( Table\,\ref{tab.spectralfit_solar}).
The spectrum is predominantly thermal, with minimal emission below 1.5 keV, showing only Si and S lines, and displaying hard spectra. They also noted its similarity to WR stars and suggested it might be a colliding-wind binary. \citet{clark08}, analysing the same data, confirmed these spectral characteristics. Their non-equilibrium plasma model fitted to the spectrum is consistent with our Model B in terms of flux and thermal temperature.  However, they measure a slightly lower ionization parameter and local absorption. 

Overall, the EWOCS data have increased the number of collected photons by a factor of 25, and extended the observational baseline to 14 months, allowing us to observe spectral features previously unseen, such as the iron line (Fe~XXV) at 6.7\,keV. The X-ray spectrum also is very similar to the EWOCS X-ray spectra of the brightest WR stars \cite[stars A, B, and L;][]{konna24}. 
Additionally, the 36 observations allow us to search and detect long-term variability (P=14 days), which was not possible with the archival data.

\subsection{Periodicity investigations for Wd1-9}

Although many studies have suggested that Wd1-9 is most likely a binary, prior to the EWOCS data, no orbital period
or clear signs of binarity in spectroscopic data had been detected, and the only direct evidence for binarity came the from X-ray spectral analysis, which found that the source exhibits a hard thermal spectrum and a high X-ray flux \citep{skinner06, clark08}.
A photometric period or clear signs of binarity in spectroscopic data in any waveband have not yet been detected. For example, \citet{bonanos07} used optical photometry covering $\sim$20 days and found that Wd1-9 is a non-periodic variable. However, the time baseline of this work makes the detection of eclipses in a $\sim$10 day system unlikely, even if they are present.
\citet{clark10} found no significant IR spectroscopic variability over four years, noting that their data resolution might not reveal the line profile variability typical of hot supergiants. Similarly, \citet{clark13} reported on a lack of spectroscopic variability over 30 years. The absence of photospheric lines in the IR spectra suggests that an envelope or cocoon obscures the binary system hindering the detection of any variability or periodicity \citep{clark05, clark10, clark13}.

We investigated potential optical periodic signals in the \gaia{} DR3 data release, where the source was flagged as variable \citep{gaiamission,gaiadr3}. Despite analysing the G, G${BP}$, and G${RP}$ bands, no periodic signals were found due to the limited number of observations (40 over $\sim$1000 days). Close monitoring is crucial for detecting any potential optical periodicity. Currently, only the X-ray EWOCS data provide a periodic measurement of $\sim$14 days.

\subsection{Nature of Wd1-9}\label{nature}

The absence of optical photospheric lines, along with the mid-IR dust and radio nebula, suggests that Wd1-9 is shrouded in a dense cocoon or envelope. This obscuration likely gives rise to the B[e] phenomenon and conceals the exact properties of the star from our view \citep{clark05,clark13,dougherty10}. Wd1-9 was suspected to be a binary with the strongest clues coming from previous \chandra{} data and the past radio eruptive behaviour which created the dense circumbinary material \citep{skinner06,clark08,dougherty10}. The EWOCS data provide compelling evidence of binarity with a detected periodic signal at $\sim$14 days and a strong iron emission line at 6.7\,keV. Such hot plasma ($>10$\,MK) is not expected in the winds of single non-magnetised massive stars, but it can be produced by a wind-wind collision zone of a binary system \citep[e.g.][]{rauw22}. 
Periodic variability in massive stars has been attributed to corotating interaction regions within the stellar wind \citep[e.g.][]{chene11,massa14}. The collective evidence, including the presence of a colliding-wind zone producing X-rays, circumbinary material, and significant radio-inferred mass loss, strongly supports a binary interpretation. This suggests that the detected period corresponds to the orbital period of the system rather than wind-driven variability.

\citet{clark13} considered the possibility that Wd1-9 is a high mass X-ray binary (HMXB). Indeed, of the known HMXB population, approximately half are supergiant systems \citep[e.g.][]{Walter2015}. While we cannot completely rule out such a scenario for Wd1-9, the great majority of supergiant HMXB population reviewed by \citet{Walter2015} have X-ray luminosities well over an order of magnitude higher than that of Wd1-9 ($L_X\sim 3\times 10^{32}$~erg~s$^{-1}$), typically in the range $10^{33}$--$10^{37}$~erg~s$^{-1}$. Moreover, HMXBs are also generally characterized by hard non-thermal X-ray spectra, whereas Wd1-9 shows persistent thermal emission with no indication of a hard non-thermal component. We therefore consider it very unlikely that Wd1-9 is a HMXB.

The possible evolutionary link between sgB[e] stars and LBVs has been discussed in various studies. Both are massive, evolved stars with strong mass loss and occupy similar regions in the HR diagram. sgB[e] stars exhibit photometric and spectroscopic variability \citep[e.g.][]{bergner95,zickgraf96,zickgraf03}, although no direct connection to LBV eruptions has been established. In particular, the radio envelope structure and the recent epoch of eruptive mass loss of Wd1-9 resemble those of LBVs \citep{dougherty10}, as does its optical photometric variability \citep[][also seen in \gaia{} data]{bonanos07}. However, as noted by \citet{clark10}, Wd1-9 currently shows no long-term spectroscopic variability, unlike LBVs and cool hypergiants. 
Moreover, the radio data \citep[e.g.][]{dougherty10} indicate that the eruptive phase ended relatively recently, within approximately a thousand years. Transitioning from such an eruption to a 14-day orbit of two massive stars in such a short time scale is not evolutionarily consistent with an LBV-driven mass-loss event. Instead, binary interaction and rapid Roche-lobe overflow is the only explanation for the  extreme mass loss rates inferred by \citet{dougherty10}.

A common envelope scenario for Wd1-9 has been discussed in previous studies \citep[i.e.][]{dougherty10,clark13}. \citet{dougherty10} noted that such a phase cannot be ruled out, as it involves significant mass loss. \citet{clark13} further supported the idea of a massive interacting binary by proposing that Wd1-9 is a low-inclination O+O or WNVL+O binary, currently undergoing or having recently completed rapid Roche-lobe overflow. This extreme mass transfer could have led to the ejection of the hydrogen envelope of the primary, forming a circumbinary disk. Theoretical models \citep[e.g.][]{wellstein99,petrovic05} suggest that such rapid Roche-lobe overflow can drive mass loss rates comparable to those inferred for Wd1-9 \citep{dougherty10}. With the detection of a 14-day period in this work and strong evidence of a colliding-wind binary, an initial configuration with unequal mass ratio would be expected to lead to common envelope evolution and merger \citep{langer03}, implying that Wd1-9 formed as an O+O binary with similar initial masses.

\subsection{A toy model}

In this section, we build on the idea presented in Section\,\ref{nature} that Wd1-9 is a binary system that has recently undergone rapid Roche-lobe overflow and is deeply embedded in a dense cocoon of circumbinary material or a disk. While the exact nature of the binary system remains unclear, previous optical and infrared studies \citep{clark05, clark13} allow us to confidently exclude certain spectral types. For instance, carbon-type (WC) primaries are excluded due to the absence of spectroscopic features such as carbon-rich dust. Although the characteristic \ion{He}{ii} and \ion{N}{iv} emission lines of early-type nitrogen (WN) Wolf–Rayet stars are not detected in the spectra,  this does not rule out their presence if these features are obscured by circumbinary material.

As discussed in \citet{clark13}, and considering the evolution of mass transfer, both WRs (WN type) and O9-B0 (super)giants could be potential current primaries. In this work, we explore these spectral types for the primary and secondary stars, assuming a Keplerian orbit with zero eccentricity and a period of 14 days.
Our aim is to investigate whether any of these combinations can reproduce the observed spectral temperature ($kT \sim 3.0$–$3.2\,\mathrm{keV}$), without yet considering their alignment with expected evolutionary pathways — this aspect is addressed in later Sections.
Throughout this work, we refer to the initially more massive star as the primary (i.e. the initial primary), regardless of its current mass.
In Table\,\ref{tab.binaryprop}, we present the adopted parameters for the possible primary and secondary stars in the binary system, using average values for their various properties based on Galactic stars of the same spectral types \citep[][]{hamann06,mokiem07,hohle10,hamann19}. 
We acknowledge that the OB templates likely do not accurately represent the secondary, as Roche-lobe overflow suggests a critically rotating mass gainer that is out of equilibrium and in an atypical state. The earliest observed stars in Wd1 are O9III \citep{clark20}, therefore the inclusion in this case of earlier-type templates aims to reflect the unusual state of the secondary.

\begin{table}[h]
\centering
\caption{Spectral type of stars used and average properties}
\begin{tabular}{lcccr}
\hline
Star & $M$& $R$ & $V_{\infty}$ & $\dot M$ \\
& $M_{\odot}$ & $R_{\odot}$ & km\,s$^{-1}$ & $M_{\odot}$ yr$^{-1}$  \\
\midrule
O7-9III& 17.00  &  14.00 & 2512 & 1.2$\times 10^{-6}$  \\
B0I& 15.00&   28.00 &1460 & 1.8$\times 10^{-6}$  \\
O9I& 24.25  &  23.20 & 1875 & 3.1$\times 10^{-6}$  \\
WN7& 28.00 &  6.30 & 1380 & 1.5$\times 10^{-5}$   \\
WN8& 41.00&  8.90  & 1120 & 6.9$\times 10^{-5}$  \\
WNVL\tablefootmark{a}& 92.00 &   85.00  & 550 & 5.0$\times 10^{-5}$\\
\hline
\end{tabular}
\tablefoot{$M$ is the mass of the star, $R$ is stellar radius, $\dot{M}$ is the mass loss rate, and $V_{\infty}$ is the terminal velocity of the wind. 
\tablefoottext{a}{The values reported here correspond to a WNVL star WR 102ka at the GC \citep[][values used here are refined by F. Najarro et al. in prep. and correspond to upper limits based on uncertainties of the extinction law]{oskinova13}.}}
\label{tab.binaryprop}
\end{table}

\subsubsection{Theoretical framework}

\noindent From Kepler's third law we can find the separation of the two stars:
\begin{equation}
    \frac{\alpha^3}{P^2} = \frac{G(M+m)}{4\pi^2},
\end{equation}

\noindent where $\alpha$ is the semi-major axis of the orbit, $P$ is the orbital period, $G$ is the gravitational constant, and $M$ and $m$ represent the masses of the primary and secondary stars, respectively.

\noindent The distance of the apex of the collision zone from each star is given by \citep{EU1993}:

\begin{equation}
    r_1=\frac{1}{1+\eta^\frac{1}{2}}\alpha,\ \  r_2=\frac{\eta^\frac{1}{2}}{1+\eta^\frac{1}{2}}\alpha,
\end{equation}
where $r_1$ and $r_2$ are the distances of the primary and the secondary star from the apex of the collision zone respectively, and $\eta$ is the so-called wind momentum rate ratio which is given by \citep{EU1993}:
\begin{equation}
    \eta = \frac{\dot{M}_{2}V_{\infty2}}{\dot{M}_1V_{\infty1}}
\end{equation}

\noindent The pre-shock velocity in the wind collision zone is given by the beta velocity law as \citep[see][]{puls2008}:

\begin{equation}
    V(r)=V_{\infty, \beta} \left(1-\frac{R_*}{r}\right)^\beta
\end{equation}
where  $r$ is the distance from the collision zone, $R_*$ the stellar radius. 
We have adopted a value of $\beta=$0.9 for O-type main sequence stars, $\beta=$1.1 for O- and B-type giant stars, and $\beta=$1.2 for WR stars based on typical values found in the literature 
\citep[i.e.][]{muijres12}.

\noindent The kinetic energy of the pre-shock gas converts into thermal energy in the post-shock region following \citep[see][]{stevens1992}:

\begin{equation}
    T=\frac{3}{16}\frac{\mu\, m_p}{k}V^2
\end{equation}
where $k$ is the Boltzmann constant, $V$ the pre-shock velocity, $\mu$ the mean molecular weight, and $m_p$ the proton mass.We adopt a mean molecular weight of 2.0 for the expanding winds of WR stars, and 1.5 for those of OB stars \citep[e.g.][]{leitherer95}.

To understand whether the shock cools adiabatically or radiatively one needs to examine the cooling time in comparison to the dynamical timescale. The cooling time $t_{\text{cool}}$ is the time it takes for the gas to radiate away its thermal energy.  For an optically-thin collision-dominated plasma that characterises colliding-wind binaries, a simplified expression for the radiative cooling time is:

\begin{equation}
    t_{\text{cool}} = \frac{3kT}{2n_e \Lambda(T)}
\end{equation}

\noindent where $k$ is the Boltzmann constant, $T$ is the temperature of the post-shock gas, $n_e$ is the electron density, and $\Lambda(T)$ is the cooling function, which depends on the temperature and metallicity of the gas. 
For the cooling function $\Lambda(T)$ we used the cooling curve from \citet{sutherland93}, and for the electron density we assumed a strong shock, where the post-shock density is four times the pre-shock density. Thus, the electron density is given by

\begin{equation}
    n_e(r) = \frac{\rho_{\mathrm{post}}}{\mu m_p} = \frac{4\rho(r)}{\mu m_p} ,
\end{equation}

\noindent where $\mu$ is the mean molecular weight, which for fully ionised post-shock winds is $\sim$1.35 for WR stars and $\sim$0.6 for OB type stars, $m_p$ is the proton mass, and $\rho(r)$ is the pre-shock wind density, given by the continuity equation

\begin{equation}\label{eq.density}
    \rho(r) = \frac{\dot{M}}{4\pi r^2 V},
\end{equation}

\noindent where $\dot{M}$ is the mass-loss rate of the star, $r$ is the distance from the star, and $V$ is the pre-shock wind velocity.

The dynamical timescale, which represents the time required for material to traverse the shock region, is given by:

\begin{equation}
t_{\text{dyn}} = \frac{r}{V}
\end{equation}

\noindent where $r$ is the distance from the star to the wind collision zone, and  
${V}$ is the velocity at shock region.  

\subsubsection{Results of the toy model}

\begin{table*}[!htbp]
\centering
\caption{Toy model results for a circular orbit}
\begin{tabular}{rllcccrrcc}
\hline
Case &Primary & Secondary & $\alpha$  & $r_1$ & $r_2$  & $V_1$  & $V_2$  & $T_1$  & $T_2$  \\ 
& & & $R_{\odot}$ & $R_{\odot}$ &  $r_{\odot}$& km\,s$^{-1}$ & km\,s$^{-1}$ & keV & keV \\ 
(1) & (2)& (3) & (4) & (5) & (6)& (7) & (8) & (9) & (10) \\ 
\hline
*1 &O7-9III & O7-9III & 79.20 & 39.60 & 39.60 & 1554 & 1554 & 7.07 & 7.07 \\
*2& B0I & O7-9III & 77.62 & 37.48 & 40.14 & 321 & 1567 & 0.30 & 7.19 \\
3  & B0I & B0I & 75.97 & 37.98 & 37.98 & 335 & 335 & 0.33 & 0.33 \\ 
*4&O9I & O7-9III & 84.47 & 45.05 & 39.42 & 845 & 1550 & 2.09 & 7.03 \\ 
5  & O9I & B0I & 83.08 & 45.73 & 37.36 & 860 & 318 & 2.17 & 0.30 \\ 
6  & O9I & O9I & 89.16 & 44.58 & 44.58 & 835 & 835 & 2.04 & 2.04 \\ 
**7&WN7 & O7-9III & 86.96 & 62.94 & 24.02 & 1215 & 960 & 5.78 & 2.70 \\ 
8  & WN7 & B0I & 85.65 & 63.15 & 22.50 & 1216 & - & 5.78 & - \\ 
**9 & WN7 & O9I & 91.40 & 63.64 & 27.76 & 1217 & 256 & 5.80 & 0.19 \\ 
*10 & WN7 & WN7 & 93.53 & 46.77 & 46.77 & 1160 & 1160 & 5.26 & 5.26 \\ 
*11 & WN7 & WN8 & 100.27 & 34.20 & 66.08 & 1080 & 941 & 4.57 & 3.46 \\  
**12&WN8 & O7-9III & 94.63 & 79.03 & 15.61 & 970 & 206 & 3.68 & 0.12 \\ 
13  & WN8 & B0I & 93.53 & 78.97 & 14.56 & 970 & - & 3.68 & - \\ 
14  & WN8 & O9I & 98.42 & 80.30 & 18.13 & 972 & - & 3.70 & - \\ 
*15  &WN8 & WN8 & 106.21 & 53.11 & 53.11 & 898 & 898 & 3.16 & 3.16 \\ 
 16 & WNVL & O7-9III & 116.78 & 87.74 & 29.05 & 8 & 1218 & - & 4.34 \\ 
17 & WNVL & B0I & 116.07 & 88.66 & 27.41 & 12 & - & - & - \\ 
18 & WNVL & O9I & 119.32 & 86.56 & 32.75 & 4 & 483 & - & 0.68 \\ 
19 & WNVL & WN7 & 120.59 & 64.57 & 56.02 & - & 1195 & - & 5.59 \\ 
20 & WNVL & WN8 & 124.79 & 46.63 & 78.17 & - & 968 & - & 3.67 \\ 
21 & WNVL & WNVL & 139.05 & 69.53 & 69.53 & - & - & - & - \\ 
\hline
\end{tabular}
\tablefoot{Columns 1–3 list the case number along with the spectral types of the primary and secondary stars. Columns 4–6 show the binary separation and the distances of each star from the apex of the wind collision zone. Columns 7 and 8 present the pre-shock velocities, while Columns 9 and 10 give the corresponding post-shock temperatures of the winds for the primary and secondary, respectively. When no values are reported for the pre-shock velocity and post-shock temperature of one of the two stars, it means that the wind from one star directly strikes the surface of the other star, preventing the development of a wind-wind collision.  Cases marked with an asterisk represent configurations that can reproduce the observed spectrum. Cases marked with a double asterisk denote physically plausible scenarios.}
\label{tab.w9toymodel}
\end{table*}

The results of the toy model are presented in Table\,\ref{tab.w9toymodel} for the various combinations of spectral types for the primary and secondary stars assuming a circular orbit. We observe that 
any combination where both stars are O or B supergiants results in thermal temperatures lower than the observed value (cases 3, 5, and 6). 
Two giant stars and a combination of an OB supergiant and an O giant can explain  the observed spectrum (cases 1, 2, and 4).
A WN7 primary with a B supergiant (case 8) can not reproduce the X-ray observations as the wind from the primary directly hits the surface of the secondary and there is no wind-wind collision.
A WN7 primary with either an O giant/supergiant or another WR star of  WN7 or WN8 type can reproduce the observed temperature (cases 7, 9, 10, and 11; $kT > 3$\,keV). 
A WN8 primary with an O or B supergiant secondary also fails to produce a colliding-wind (cases 13 and 14), while a system consisting of two WN8 stars or a WN8 and an O giant star could reproduce the observed spectrum (cases 12 and 15). A system featuring a later-type WR star (WNVL; cases 16 to 21) fails to replicate the observed X-ray spectrum due to its lower post-shock temperatures and/or non-existent wind-wind interaction.

We examined all cases (marked with an asterisk) in Table \ref{tab.w9toymodel} that represent plausible scenarios for a binary system where a wind-wind collision is possible, and found that the radiative cooling times range from 15 minutes to two hours. In contrast, the dynamical timescale spans from 6 to 14 hours. The fact that $t_\text{cool} < t_\text{dyn}$
in all these cases suggests that the shocked material cools radiatively. As it cools, the gas undergoes compression, leading to the formation of a denser region.

Additionally, EWOCS data indicate that while the average absorbed X-ray luminosity is $ L_X = 5.0 \times 10^{32} \text{ erg s}^{-1}$, variations in count rate from a minimum of 0.005 ct/s (corresponding to $L_{\text{min}} = 1.91 \times 10^{32} \text{ erg s}^{-1}$, $f_{\text{min}} = 8.94 \times 10^{-14} \text{ erg cm}^{-2} \text{s}^{-1}$) to a maximum of 0.105 ct/s ($L_{\text{max}} = 4.24 \times 10^{33} \text{ erg s}^{-1}$, $f_{\text{max}} = 1.98 \times 10^{-12} \text{ erg cm}^{-2} \text{s}^{-1}$) suggest significant flux variability\footnote{The count rate to flux conversion was performed using model spectrum A as a baseline, adjusting only the normalisations of the thermal components.}.
If we assume that these flux variations are not due to absorption effects, as \nh{} remains relatively constant within uncertainties (Fig.\,\ref{fig.colorw9}) for the different observations, then the X-ray variability likely arises from changes in the density of shocked wind material along an eccentric orbit. 
From the observed flux ratio, we estimate the density at maximum emission to be $\sim 4.56$ times higher than at minimum. Given its inverse-square dependence on separation (Eq.\,\ref{eq.density}), this implies an orbital eccentricity of $e \approx 0.36$. It is important to note that this represents an upper limit on eccentricity as flux variations may arise from factors beyond changes in density along an eccentric orbit. However, as discussed in Section\,\ref{nature}, the system has likely just emerged from a rapid Roche-lobe overflow phase, which would have generally circularised the orbit or left it with very low eccentricity, even if the pre-interaction system had a high eccentricity. We therefore exclude this scenario, as it does not align with the expected evolutionary history of the system.
For completeness, we briefly present these results in the appendix (Section \ref{eccentric}).

Overall, it appears that the presence of a WR star (WN7 or WN8), or a giant O star (O7-9III) is necessary to reproduce the X-ray observations (Table\,\ref{tab.spectralfit_solar}) while given the evolutionary state of the binary a high eccentricity orbit is unlikely. We consider WN7 as representative for H-free WN stars, although earlier sub-types are possible and would also explain the observations.
It is important to note that the values obtained from the toy model are only approximate, as several uncertainties are introduced. For example, all stellar and wind properties used in this analysis are based on average values of Galactic stars and are compared to the mean spectral properties of Wd1-9 over a period of approximately one year. Moreover, the template used for the WNVL star in Table\,\ref{tab.binaryprop} corresponds to a Galactic Centre star, likely with a higher metallicity than that of Wd1. However, even when using average values based on LMC WR stars from \citep{crowther97}, which exhibit lower metallicity than Wd1, our main conclusions remain unchanged.
Additionally, a colliding-wind scenario could still develop in cases where our toy model suggests no wind-wind collision, if mutual radiative effects slow down the primary wind and prevent it from directly impacting the surface of the secondary \citep{stevens1994,gayley1997}, although with lower pre-shock velocity for the primary star. 
Therefore, while a WR primary remains the most plausible scenario, confirming this will require further observational or theoretical support. In this context, upcoming EWOCS work will  will address this by comparing hydrodynamic models of outflows from various binary configurations with ALMA and JWST observations.

\subsection{Wd1-9 placed in  evolutionary context}

Based on the results of our toy model and the fact that we do not expect an eccentric orbit after a rapid Roche-lobe overflow phase, the most likely scenario is that Wd1-9 hosts a WN7/WN8  or an O7-9III star (Table\,\ref{tab.w9toymodel}). Key spectral features characteristic of the WN types, such as the \ion{N}{iv} emission line, are absent in the optical spectra \citep{clark13}. This absence could indicate a cooler spectral type, such as WN9-11h, however this is inconsistent with our modelling further supporting that the primary has lost its hydrogen envelope. Alternatively, the expected \ion{N}{iv} features from a WN7/WN8 primary may be entirely obscured by the dense circumbinary cocoon or disk. With an estimated size of ~800 AU, an inclination of ~60 degrees, and likely puffed up by radiation pressure \citep{clark13}, this structure could completely enshroud both stellar components. As a result, the observed optical spectrum may be dominated by stationary disk emission rather than Doppler-shifted stellar lines. Thus, while the spectral features remain hidden, the presence of a WN7/WN8 primary, whose powerful winds produce the observed X-ray emission, is strongly supported by our modelling and remains among the most likely interpretations.

Wd1-9 exhibits several key observational characteristics indicative of a post-interaction system. Significant mass loss and circumbinary material suggest a history of rapid mass transfer. This is further supported by the discovery of a 14-day orbital period in this work.
While post-Roche lobe overflow systems can, in principle, evolve into configurations with components like O7–9III + B0I or O9I, such pairings appear to be uncommon in observed systems (see cases 2 and 4 in Table\,\ref{tab.w9toymodel}), particularly following a rapid mass transfer phase.
Eruptive behaviour typically points to unstable or non-conservative mass transfer, often accompanied by significant mass and angular momentum loss, which tends to strip the donor star down to its helium core. For a system to retain a B0I or O9I donor, the mass transfer must have been only partially unstable, allowing the star to hold onto enough of its hydrogen envelope to avoid becoming a Wolf-Rayet. Similarly, forming an O7-9III + O7-9III system (case 1 in Table\,\ref{tab.w9toymodel})  would require highly fine-tuned conditions—such as nearly equal initial masses and stable, late-stage mass exchange—which are rare, especially in systems  with an eruptive history.

The presence of a primary that has almost completely lost its hydrogen envelope (WN7/WN8 star) suggests that interaction took place while the donor was in the shell hydrogen-burning phase, placing the system within the early Case B mass transfer regime \citep{wellstein99,petrovic05}.
Compared to Case A, which involves mass transfer near the end of core hydrogen burning and progresses more gradually, Case B leads to a more rapid increase in helium abundance and a corresponding decrease in hydrogen. This faster mass-loss phase naturally leads to the formation of hydrogen-poor WN7/WN8 stars.
In Case B, mass transfer begins after the donor has exhausted core hydrogen and expands as it burns hydrogen in a shell. Early Case B interactions occur before the donor evolves into a fully developed supergiant, leading to more efficient stripping of the outer envelope without immediately triggering a common envelope phase. This aligns well with Wd1-9, where the donor has already lost much of its hydrogen envelope (assuming is a WN7/WN8 star) but remains in a binary system rather than having merged.
Meanwhile, following rapid mass transfer (early case B), the secondary star is expected to be a rapidly rotating, hydrogen-rich, underluminous OB star.
In this scenario, and  up to the current evolutionary stage, the formation of a second WR star appears unlikely, as producing a WN companion would require reverse mass transfer, which contradicts the presence of the observed circumbinary material. Consequently, Cases 7, 9, 12 in Table\,\ref{tab.w9toymodel} (i.e. WN7/WN8 + O9I/O7-9III) emerge as the most physically plausible configurations.

Wd1-9, having just exited an early Case B mass transfer, represents a rare transitional system bridging pre- and post-interaction massive binaries in Westerlund 1. A possible precursor analogue is Wd1-24, a system of two O9 supergiants in a 6.5-day orbit, where mass transfer has yet to begin \citep{ritchie22}. While other immediate precursor analogues likely exist, their brief mass transfer phases make them difficult to observe, even in a massive cluster like Westerlund 1.  
More evolved post-interaction systems include Wd1-13 (WNVL+OB), which exhibits non-conservative mass transfer, with the donor at $\sim$23~$M_{\odot}$ and the gainer at $\sim$35~$M_{\odot}$ \citep{ritchie10}. Although it is not an exact match to Wd1-9 due to its shorter orbital period ($\sim$9 days), Wd1-13 provides insight into the potential evolution of Wd1-9. Other long-period post-interaction WR binaries in Wd1 include star L (Wd1-44), where the mass gainer is likely embedded in an accretion disk, and star A (Wd1-72) \citep{ritchie22, konna24}. Notably, the EWOCS X-ray spectrum of Wd1-9 closely resembles that of stars A and L, both among the brightest WR stars in Wd1 with WN7 and later spectral types \citep{konna24}.  
Infrared and X-ray studies suggest a near-100\% binarity rate among WR stars in Wd1 \citep{crowther06, konna24}, indicating that many of the 24 WR binaries in the cluster could represent the future state of Wd1-9. An example is the even more evolved post-interaction system, star S (Wd1-5), a post-supernova binary remnant \citep{clark14, ritchie22}.

Westerlund 1 hosts a rich population of evolved massive stars, including red supergiants, yellow hypergiants, and other cool-phase stars \citep[e.g.][]{clark20}. Given their large sizes, any binary companions would need to have extremely long orbital periods---on the order of many years---effectively making these stars appear isolated. This distinction highlights the possibly different evolutionary pathways for massive stars: while binary systems, such as Wd1-9, often undergo mass transfer and evolve into WR stars, single stars or those in very wide binaries are more likely to evolve into cool-phase supergiants.

\section{Conclusions}\label{conclusions}

In this study, we utilised $\sim$1\,Ms \chandra{} observations from the Extended Westerlund 1 and 2 Open Clusters Survey (EWOCS) to investigate the nature of the  sgB[e] star Wd1-9 in Westerlund 1.
Wd1-9 is one of only two known examples of sgB[e] stars found within star clusters, providing a unique opportunity to place this rare phase of massive stars into evolutionary context.
The star is considered to be a binary system hidden within a cocoon of dust with the strongest hints of binarity in the pre-EWOCS era coming from X-ray and radio data. However, since the star is deeply embedded in thick dust, its nature remained puzzling, with no clear signs of binarity or a detected period.
Leveraging the depth and long baseline of the EWOCS dataset, we present the following findings:

\begin{itemize}
    \item We detect for the first time a periodic signal of 14 days, that we interpret as the orbital period.
    \item Similar to previous \chandra{} studies, we observe a hard thermal X-ray spectrum ($\sim$3\,keV). However, this time we detect strong emission lines from various elements (Si, S, Ar) and, notably, for the first time, the iron emission line at 6.7\,keV—a clear signature of binarity.
    \item The spectrum closely resembles those of bright WR binaries in Westerlund 1, such as stars A, B, and L, which have periods ranging from 3.5 to 81 days and spectral types WN7-9.
    \item Examining Wd1-9 in X-ray colour-colour diagrams (Fig.\,\ref{fig.colorw9}), we observe variations in thermal temperature across different EWOCS observations.
    \item Folding the colours in phase, assuming a 14-day period, reveals that soft emission peaks at slightly later phases than hard emission and overall the normalisation of the soft component is higher than that of the hard component.
    \item Analysing the combined event file as a function of energy and phase, we find that at phase 0.35, where the broad and hard emission peaks, nearly all elements (except Ar) peak, including iron K~$\alpha$ fluorescent emission at 6.4\,keV, while the element Si peaks at later phases ($\sim$0.5) and appears to be driving the flux peak observed for soft energies at this phase. 
    \item A fourteen day orbital period suggests the binary system recently underwent an early Case B mass-transfer scenario. 
    \item A toy model suggests a system consisting of WN7 WR donor and an underluminous OB mass gainer can reproduce the observed X-ray spectrum.

\end{itemize}

We conclude that Wd1-9 is most likely classified as SgrB[e] due to the optical emission originating entirely from its dense circumbinary material, making this designation largely phenomenological.
Building on previous optical and infrared studies, as well as the detection of a 14-day periodic signal and the strong 6.7\,keV iron emission line in the EWOCS data, Wd1-9 appears to be a binary system (WN7+OB) that has recently undergone an early-case B mass transfer scenario.
This places it in a rare evolutionary stage, having just undergone rapid Roche-lobe overflow and probably revealing a newly born WR star. As a result, Wd1-9 serves as a crucial link between an immediate progenitor analogue such as Wd1-24, and descendent systems, including long-period WR stars, such as stars A and L, as well as post-supernova systems like Wd1-5 within Westerlund 1.

\begin{acknowledgements}
KA acknowledges support from \chandra{} grants GO3-24033B, GO0-21010X, and TM9-20001X, and \textit{JWST} grant JWST-GO-01905.002-A. IN is partially supported by the Spanish Government Ministerio de Ciencia e Innovaci\'on (MCIN) and Agencia Estatal de Investigaci\'on (MCIN/AEI/10.130~39/501~100~011~033/FEDER, UE) under grant PID2021-122397NB-C22, and also by MCIN with funding from the European Union NextGenerationEU and Generalitat Valenciana in the call Programa de Planes Complementarios de I+D+i (PRTR 2022), project HIAMAS, reference ASFAE/2022/017.
CJKL gratefully acknowledges support from the International Max Planck Research School for Astronomy and Cosmic Physics at the University of Heidelberg in the form of an IMPRS PhD fellowship.
JFAC acknowledge  support from the CONICET research council and the University of Rio Negro, Argentina.
AB acknowledges support from the Deutsche Forschungsgemeinschaft (DFG, German Research Foundation) under Germany's Excellence Strategy – EXC 2094 – 390783311
The scientific results reported in this article are based on observations made by the \chandra{} X-ray Observatory and data obtained from the \chandra{} Data Archive.
SS acknowledges funding from the European Union under the grant ERC-2022-AdG, "StarDance: the non-canonical evolution of stars in clusters", Grant Agreement 101093572, PI: E. Pancino. 
This research has made use of software provided by the \chandra{} X-ray Center (CXC) in the application packages CIAO and Sherpa.
This work has made use of data from the European Space Agency (ESA) mission
{\it Gaia} (\url{https://www.cosmos.esa.int/gaia}), processed by the {\it Gaia}
Data Processing and Analysis Consortium (DPAC,
\url{https://www.cosmos.esa.int/web/gaia/dpac/consortium}). Funding for the DPAC
has been provided by national institutions, in particular the institutions
participating in the {\it Gaia} Multilateral Agreement.
We also made use of NASA’s Astrophysics Data System.
\end{acknowledgements}

%



\bibliographystyle{aa} 
\bibliography{references} 



\begin{appendix} 
\onecolumn
\section{Phase-binned spectra}

\begin{figure*}[htbp]
       \includegraphics[width=0.4\columnwidth]{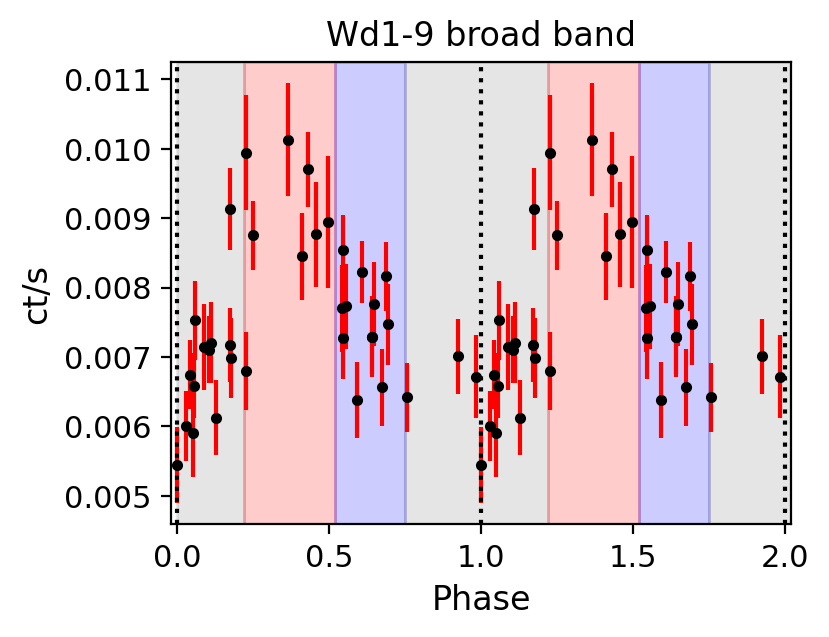}
 
   \caption{Phase-folded light curve of Wd1-9 and bin selection. The three coloured regions (grey, red, blue) indicate the phase bins used for spectral extraction.}
   		\label{fig.lcbroadbin}
   \end{figure*}

\begin{figure*}[htbp]
   	\centering	
    \includegraphics[width=0.33\columnwidth]{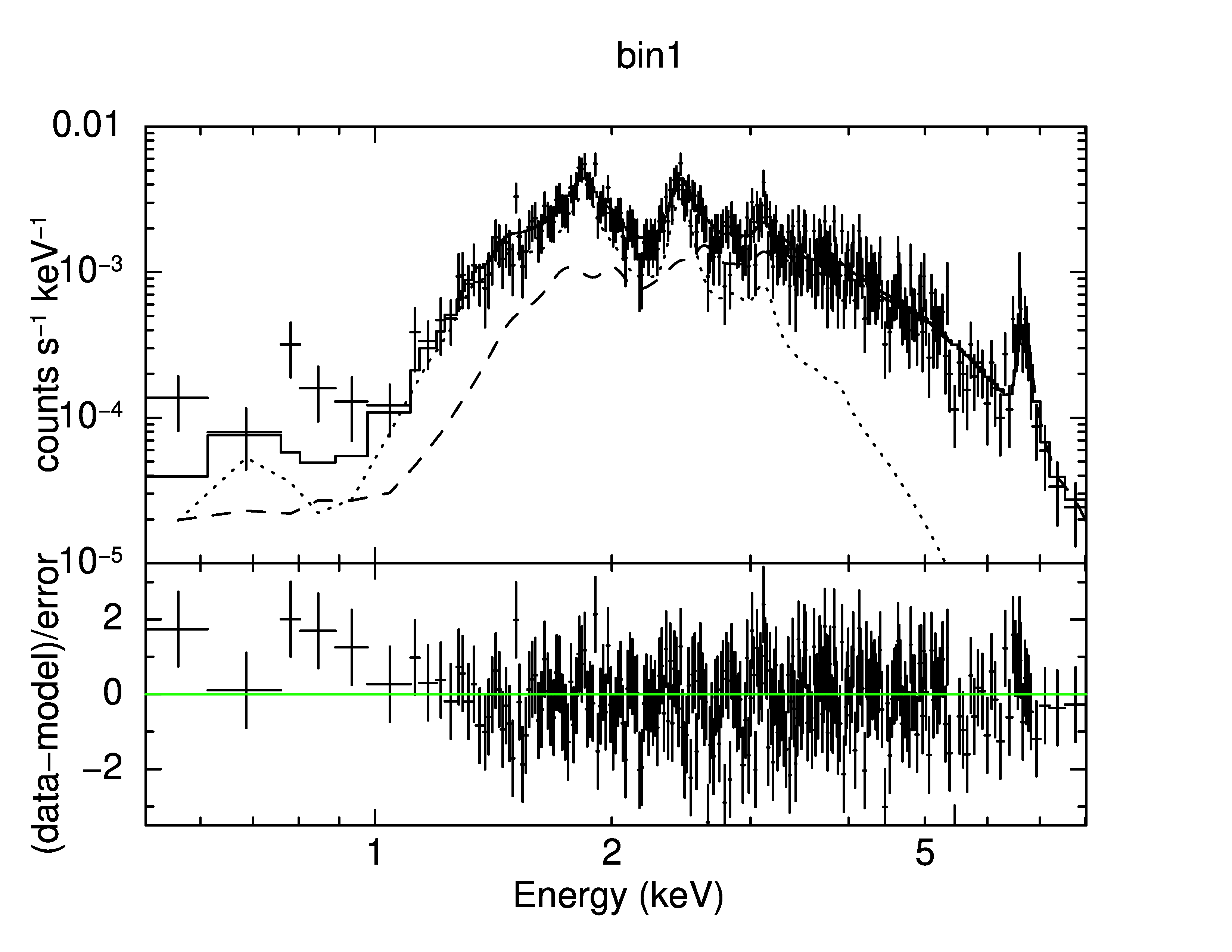}
   \includegraphics[width=0.33\columnwidth]{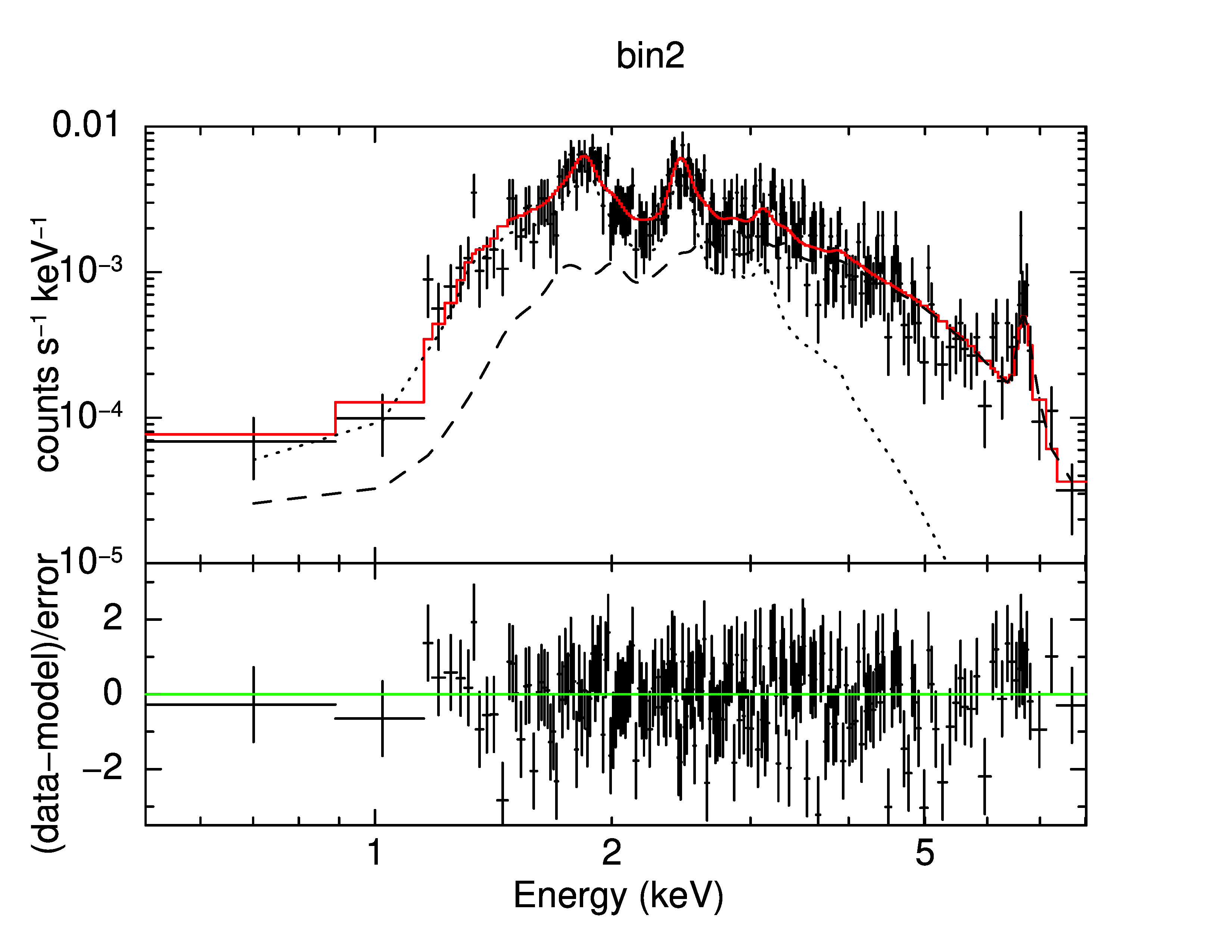}
    \includegraphics[width=0.33\columnwidth]{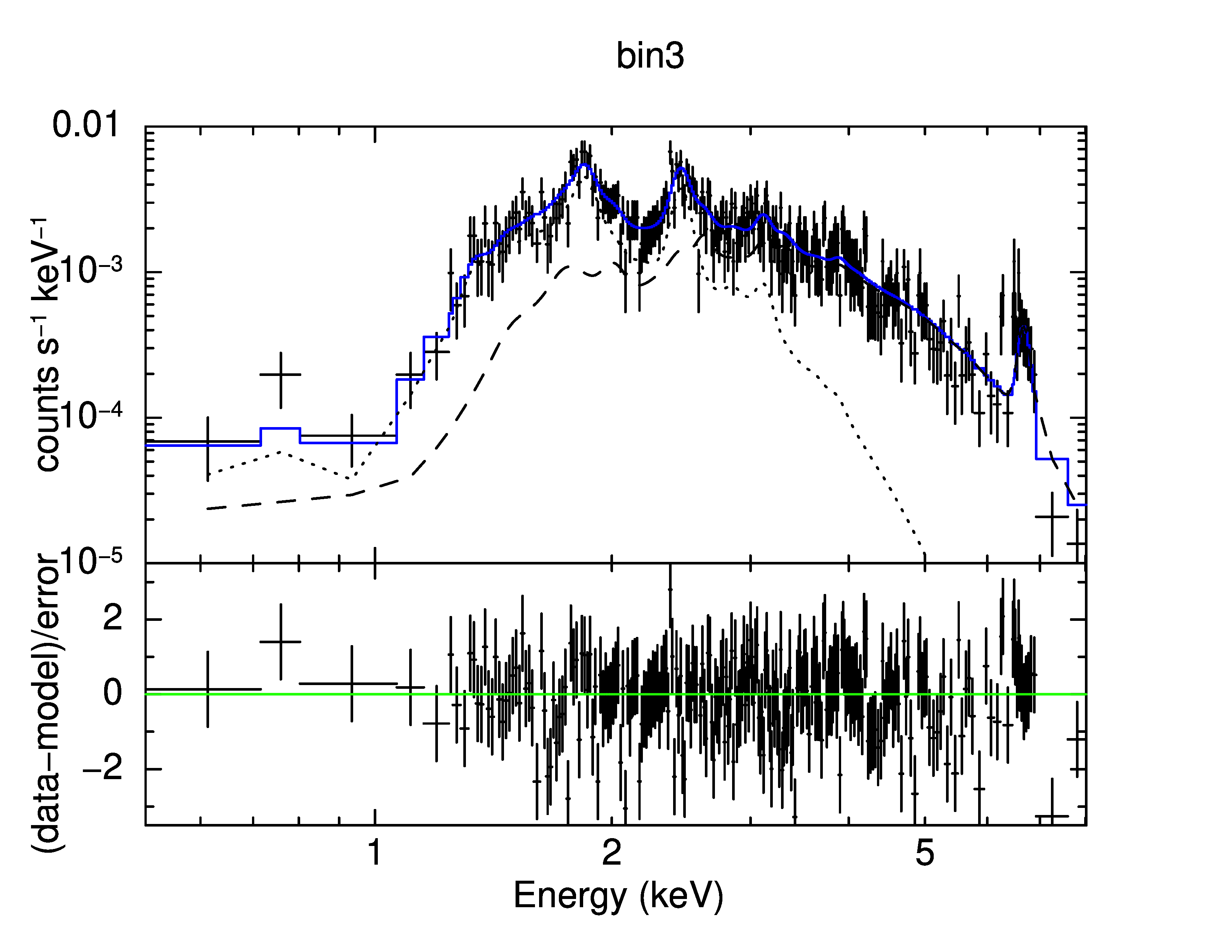}
   \caption{Spectral modelling of Wd1-9 in the three phase bins using Model A.}
   		\label{fig.phasedspectra}
   \end{figure*}

In this section, we examine the spectral properties of the phase-binned spectra of Wd1-9 using only the EWOCS observations, aiming to identify any potential variations.
Based on the light curve count rate, we defined three broad phase bins: bin 1 (grey), covering phases 0–0.22 and 0.75–1.0; bin 2 (red), covering phases 0.22–0.52; and bin 3 (blue), covering phases 0.55–0.75 (Fig.\,\ref{fig.lcbroadbin}). For simplicity, we adopted Model A and fixed the abundances to the best-fit values from the combined spectrum. As shown in Table\,\ref{tab.spectralfit_phased}, although there is some variation in absorption and in the temperatures of the two thermal components across phase, the differences are not statistically significant, as all values are consistent within the uncertainties. However, the soft excess observed in the combined spectrum appears more pronounced in bin 1 (left panel of Fig.\,\ref{fig.phasedspectra}), which also corresponds to the phase interval with the lowest absorption. This behaviour is reminiscent of the cool component reported in WR\,140 \citep{sugawara15}.

\begin{table*}[htbp]
 	\centering
 		\caption{X-ray spectral fitting results for the phased-bin spectra}
             \begin{tabular}{@{}lllllcc@{}}
			\toprule		
 		Bin &  $N\rm{_{H}^{local}}$ & $kT$ & norm & Prob ($\chi^2$/dof)  & $F_X$ &  $L_X$ \\[2pt]
 &  $10^{22}$\,\cmsq  &  keV  & cm$^{-5}$ &&  10$^{-13}$\funit & 10$^{32}$\ergs\\
 \midrule
 1 (grey) & 3.01$_{-0.44}^{+0.65}$ &0.65$\pm0.14$ &(0.96$_{-0.34}^{+1.08}$)e$-$3  &0.32 (271.60/273)& 1.40$_{-0.23}^{+0.03}$ &3.02$_{-0.50}^{+0.04}$ \\ [3pt]
  &  &  3.16$_{-0.53}^{+0.77}$ &(1.94$_{-0.42}^{+0.57}$)e$-$4  &&& \\ [3pt]
 2 (red) & 3.46$_{-0.68}^{+0.90}$ &0.59$\pm0.14$ &(2.02$_{-0.99}^{+3.50}$)e$-$3  &0.22 (197.93/202)& 1.78$_{-0.58}^{+0.04}$ &3.81$_{-1.31}^{+0.09}$\\ [3pt]
  &  &  3.23$_{-0.66}^{+1.07}$ &(2.35$_{-0.65}^{+0.89}$)e$-$4  &&& \\ [3pt]
  3 (blue) & 3.58$_{-0.58}^{+0.65}$ &0.54$\pm0.11$ &(2.15$_{-1.00}^{+2.47}$)e$-$3  &0.27 (279.84/253)& 1.53$_{-0.37}^{+0.03}$ &3.27$_{-0.79}^{+0.07}$ \\ [3pt]
  &  &  2.75$_{-0.37}^{+0.63}$ &(2.45$_{-0.58}^{+0.60}$)e$-$4  &&& \\ [3pt]
 \bottomrule
     \end{tabular}
     \tablefoot{X-ray spectral fitting results for the phase-binned EWOCS spectra using Model A. Abundances are fixed to the best-fit values from the combined spectrum. The fits were performed on spectra grouped to a minimum of five counts per bin using Cash statistics.} 
    \label{tab.spectralfit_phased}
 \end{table*}

  \begin{figure*}[htbp]
    	\centering	
    \includegraphics[width=0.33\columnwidth]{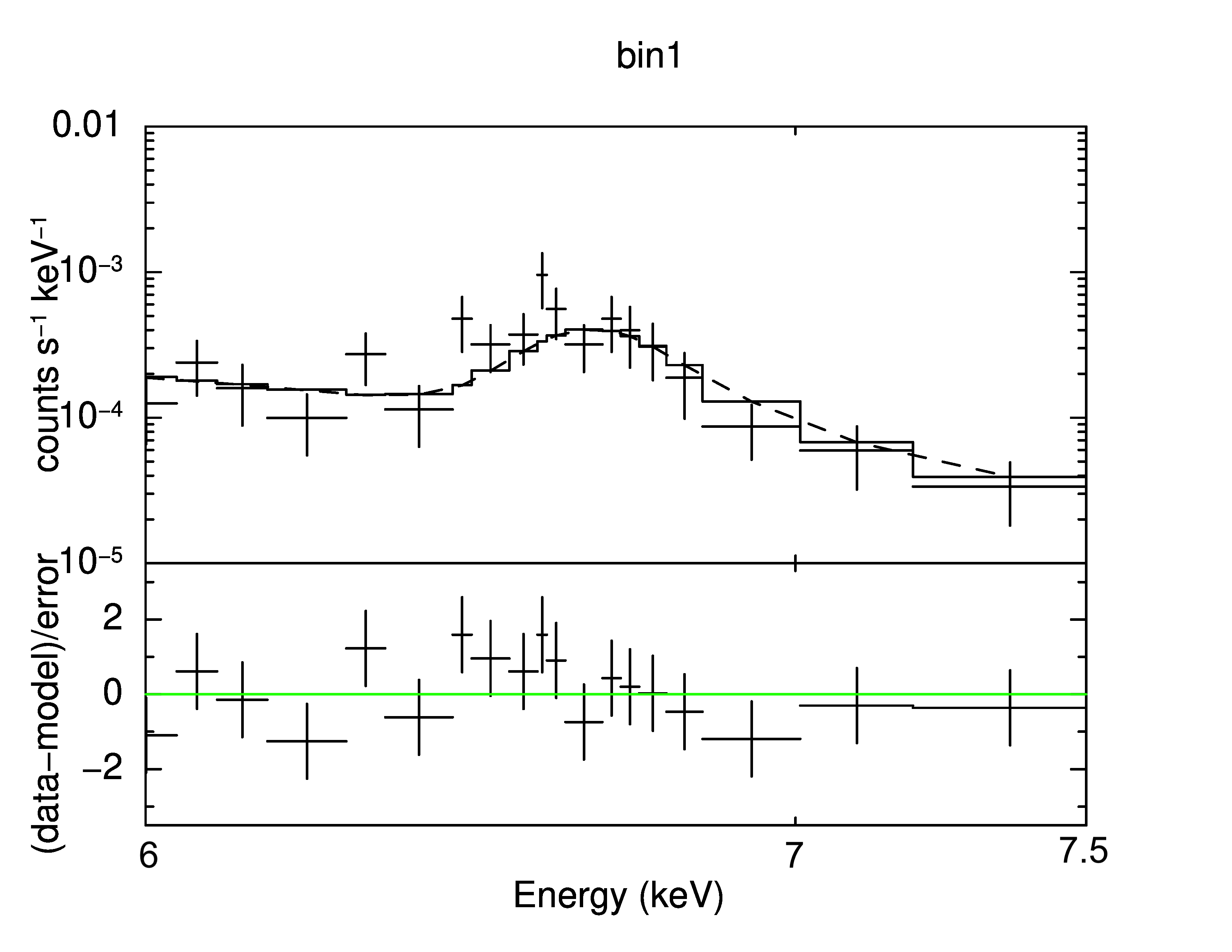}
   \includegraphics[width=0.33\columnwidth]{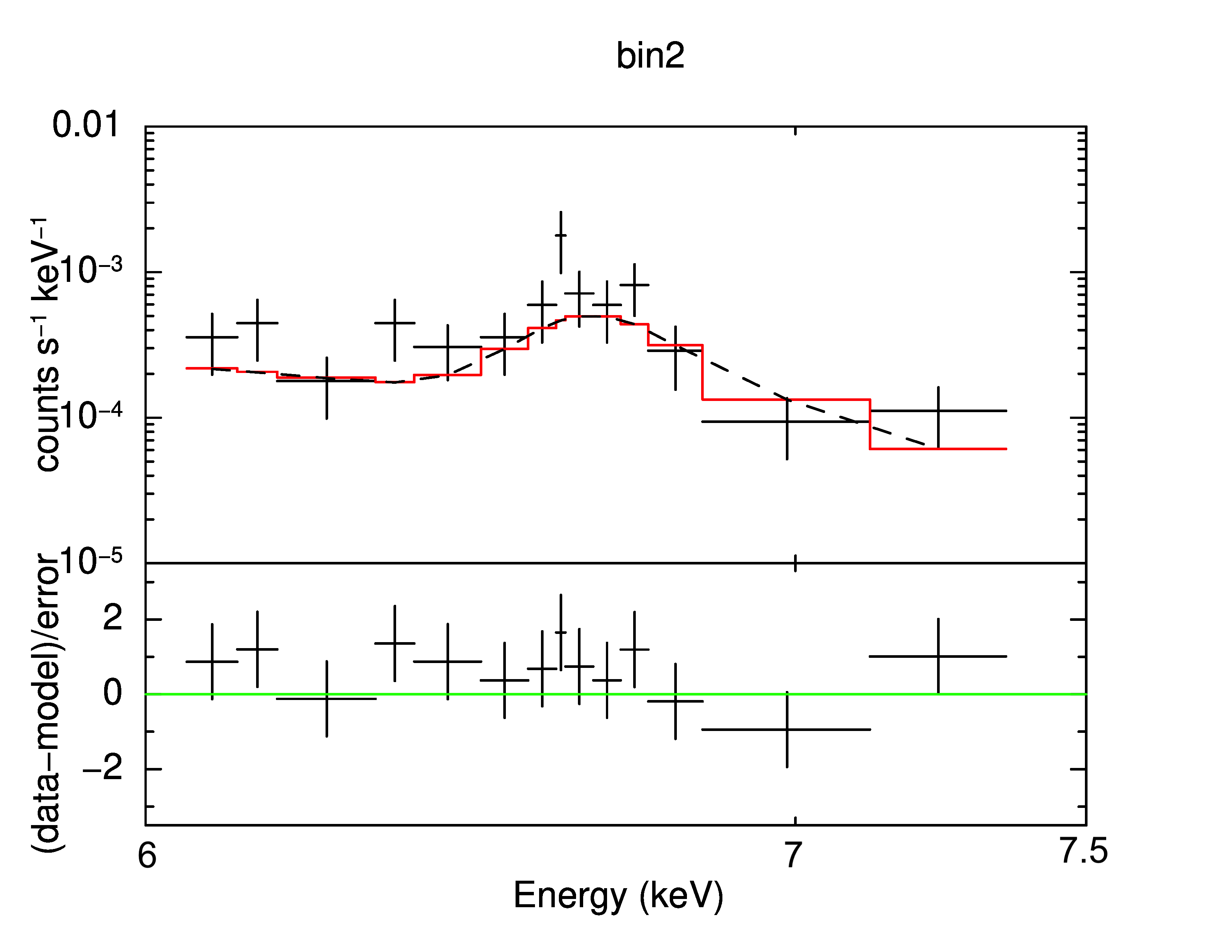}
    \includegraphics[width=0.33\columnwidth]{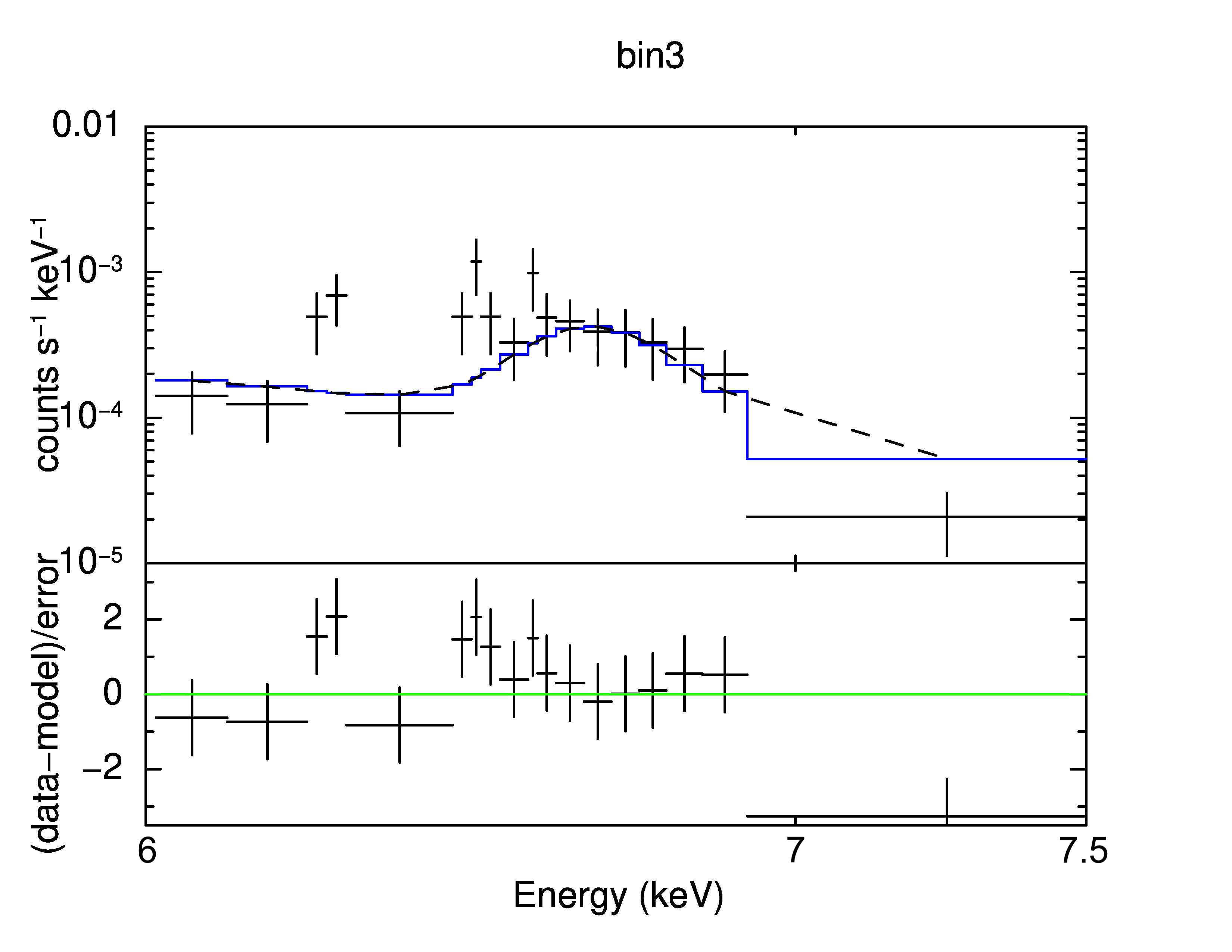}
       \caption{Same as Fig.\,\ref{fig.phasedspectra} but zoom-in on the iron line region for Wd1-9.}
   		\label{fig.phasedspectrazoom}
  \end{figure*}

\FloatBarrier

 \section{Toy model results for an eccentric orbit}\label{eccentric}

In this section we present the toy model results (Table\,\ref{tab.eccentricity}) for an eccentric orbit. While such models can reproduce the observed spectrum, they are considered highly unlikely and inconsistent with the expected evolutionary scenario.

\begin{table*}[!h]
\centering
\caption{Results of the toy model for an orbit with eccentricity e=0.36 }
\setlength{\tabcolsep}{3pt} 
\begin{tabular}{cllccccccrrccrrcc}
\hline
Case &Primary &Secondary & $\alpha_{per}$ & $\alpha_{ap}$& $r_{1per}$ & $r_{2per}$ &$r_{1ap}$ & $r_{2ap}$ & $V_{1per}$  & $V_{2per}$  & $T_{1per}$  & $T_{2per}$   & $V_{1ap}$  & $V_{2ap}$ & $T_{1ap}$   & $T_{2ap}$  \\
&& & $R_{\odot}$ & $R_{\odot}$ & $R_{\odot}$ & $R_{\odot}$ & $R_{\odot}$& $R_{\odot}$&  km\,s$^{-1}$ &  km\,s$^{-1}$  & keV  & keV  &  km\,s$^{-1}$  &  km\,s$^{-1}$  & keV  & keV  \\
(1)&(2)& (3) & (4) & (5) & (6) & (7)& (8)&  (9) & (10)  & (11)  & (12) & (13)  &  (14) & (15) & (16) &(17) \\
\hline
*1&O7-9III & O7-9III & 50.69 & 107.71 & 25.34 & 25.34 & 53.86 & 53.86 & 1037 & 1037 & 3.15 & 3.15 & 1803 & 1803 & 9.52 & 9.52 \\
*2&B0I & O7-9III & 49.67 & 105.56 & 23.99 & 25.69 & 50.97 & 54.59 & - & 1056 & - & 3.27 & 607& 1813 & 1.08 & 9.62 \\
3&B0I & B0I & 48.62 & 103.31 & 24.31 & 24.31 & 51.66 & 51.66 & - & - & - & - & 618 & 618 & 1.12 & 1.12 \\
*4 &O9I & O7-9III & 54.06 & 114.88 & 28.83 & 25.23 & 61.27 & 53.61 & 311 & 1031& 0.28 & 3.11 & 1110 & 1800 & 3.61 & 9.49 \\
*5&O9I & B0I & 53.17 & 112.99 & 29.27 & 23.91 & 62.19 & 50.81 & 332& - & 0.32 & - & 1121 & 604 & 3.68 & 1.07 \\
*6 & O9I & O9I & 57.06 & 121.25 & 28.53 & 28.53 & 60.63 & 60.63 & 296& 296 & 0.26 & 0.26 & 1102& 1102 & 3.56 & 3.56 \\
*7&WN7 & O7-9III & 55.65 & 118.26 & 40.28 & 15.37 & 85.60 & 32.66 & 1125 & 176 & 4.95 & 0.09 & 1259 & 1357 & 6.20 & 5.39 \\
*8 &WN7 & B0I & 54.82 & 116.48 & 40.42 & 14.40 & 85.88 & 30.60 & 1126 & - & 4.96 & - & 1259 & 96 & 6.20 & 0.03 \\
*9 &WN7 & O9I & 58.49 & 124.30 & 40.73 & 17.76 & 86.55 & 37.75 & 1128 & - & 4.97 & - & 1260 & 656 & 6.21 & 1.26 \\
*10 &WN7 & WN7 & 59.86 & 127.21 & 29.93 & 29.93 & 63.60 & 63.60 & 1039 & 1039 & 4.22 & 4.22 & 1217 & 1217 & 5.80 & 5.80 \\
*11 &WN7 & WN8 & 64.18 & 136.37 & 21.89 & 42.29 & 46.51 & 89.86 & 918 & 843 & 3.30 & 2.78 & 1158 & 988 & 5.25 & 3.82 \\
*12&WN8 & O7-9III & 60.57 & 128.70 & 50.58 & 9.99 & 107.48 & 21.23 & 887 & - & 3.08 & - & 1009 & 767& 3.98 & 1.72 \\
13 &WN8 & B0I & 59.86 & 127.21 & 50.54 & 9.32 & 107.40 & 19.81 & 887 & - & 3.08 & - & 1009 & - & 3.98 & - \\
*14 &WN8 & O9I & 62.99 & 133.86 & 51.39 & 11.60 & 109.21 & 24.65 & 891 & - & 3.11 & - & 1011 & 83 & 4.00 & 0.02 \\
*15&WN8 & WN8 & 67.98 & 144.45 & 33.99 & 33.99 & 72.22 & 72.22 & 778 & 778 & 2.37 & 2.37 & 956 & 956 & 3.58 & 3.58 \\
*16 &WNVL & O7-9III & 74.74 & 158.83 & 56.15 & 18.59 & 119.32 & 39.50 & - & 539 & - & 0.85 & 123& 1552 & 0.06 & 7.05 \\
17 &WNVL & B0I & 74.28 & 157.85 & 56.74 & 17.54 & 120.57 & 37.27 & - & - & - & - & 127 & 316 & 0.06 & 0.29 \\
18 & WNVL & O9I & 76.36 & 162.27 & 55.40 & 20.96 & 117.73 & 44.55 & - & - & - & - & 118 & 834 & 0.05 & 2.04 \\
19 &WNVL & WN7 & 77.18 & 164.00 & 41.32 & 35.85 & 87.81 & 76.19 & - & 1094 & - & 4.68 & 9 & 1244 & - & 6.05 \\
20 &WNVL & WN8 & 79.87 & 169.72 & 29.84 & 50.03 & 63.41 & 106.30 & - & 885 & - & 3.06 & - & 1008 & - & 3.98 \\
21 &WNVL & WNVL & 88.99 & 189.11 & 44.50 & 44.50 & 94.56 & 94.56 & - & - & - & - & 35 & 35 & - & - \\
\hline
\end{tabular}
\tablefoot{The notation follows that of Table\,\ref{tab.w9toymodel} with subscripts "$ap$" and "$per$" denoting values at apastron and periastron, respectively. Cases marked with an asterisk represent configurations that can reproduce the observed spectrum.}
\label{tab.eccentricity}
\end{table*}

We find that the models marked with an asterisk can reproduce the observed values. However, in some cases (1, 2, 4, 5, and 6), we observe significant variations in post-shock temperatures between apastron and periastron that are not supported by the observational data in Fig. \ref{fig.colorw9}, or even in other instances (cases 2, 5, 8, 9, 12, 14, and 16), a wind-wind collision cannot develop during periastron.
For the sake of completeness, the scenario of a wind-photosphere collision could occur especially in the framework of the most asymmetric cases we considered. This may typically happen when dealing with a WR primary whose strong wind completely overwhelms that of a late O-type main-sequence star or an early B-type secondary. In such a scenario, the primary wind produces a shock close to the surface of the secondary, and the post-shock plasma is expected to produce thermal X-ray emission. According to \citet{usov1992} (their Eq.\,81), the thermal X-ray emission can be predicted knowing the separation between the stars, the wind parameters of the primary and the radius of the secondary. With the typical numbers quoted in Table\,\ref{tab.binaryprop} and Table\,\ref{tab.w9toymodel}, we derive X-ray luminosities well above 10$^{33}$\,erg\,s$^{-1}$. Even though this relation only accounts for bremsstrahlung (without emission lines), this number is clearly in excess of the value we observed (see Table\,\ref{tab.spectralfit_solar}). However, as discussed by \citet{arora2024} for a similar case, the radiative nature of the shock is very likely to reduce the efficiency of the conversion of wind kinetic power to thermal X-ray luminosity. This is due to the development of the thin-shell instability in the shock \citep{vishniac1994,walder1998}, leading to a somewhat corrugated structure of the shock front. This has a very strong impact on the obliquity of the wind flow on the shock surface, drastically reducing the component of the wind velocity normal to the shock front. The thin-shell instability in radiative shocks can lead to a drop of the X-ray luminosity by a factor of a few tens \citep{kee2014,steinberg18}. In the case of the short-period, asymmetric O-type binary HD\,93205, \citet{arora2024} suggested that a wind-photosphere collision is likely to occur at least in a significant part of the eccentric orbit, leading to a potential hybrid case switching between a wind-wind and a wind-photosphere interaction as a function of the orbital phase.

We observe that higher eccentricity leads to higher temperatures at apastron compared to those in the circular orbit (see Tables\,\ref{tab.w9toymodel} and \ref{tab.eccentricity}) and as expected, plasma temperatures are higher at apastron compared to periastron due to higher pre-shock velocities.
Given our assumptions, higher wind densities at periastron should result in increased flux, while lower pre-shock velocities yield cooler (softer) heated gas. This aligns with our observations (Fig.\,\ref{fig.phasedlcnorm}), where the softer emission shows significantly higher flux, while the harder emission remains generally weaker. However, the phase difference we see between the peaks of soft and hard emission (see Fig.\,\ref{fig.phasedlcnorm}) could arise from various other factors.
Detailed modelling is needed to disentangle these effects in the case of an eccentric orbit.

\end{appendix}

\end{document}